\documentclass[conference]{IEEEtran}
\IEEEoverridecommandlockouts
  
\usepackage{graphics} 
\usepackage{subcaption}
\usepackage{tabularx}
\usepackage{multirow}
\usepackage[cmex10]{amsmath}
\usepackage{amssymb, mathtools}
\usepackage{algorithm}
\usepackage{algpseudocode}
\usepackage{color}

\usepackage{enumitem}
\setlist{itemsep=0pt,labelindent=0pt,leftmargin=*}

\usepackage{url}

\algblock{Input}{EndInput}
\algnotext{EndInput}
\algblock{Output}{EndOutput}
\algnotext{EndOutput}
 
\makeatletter
\algrenewcommand\ALG@beginalgorithmic{\footnotesize}
\makeatother

\newif\iflongversion
\longversionfalse

\newcommand{\squeezeuppicture}{} %reduce whitespace after pictures

\newcommand{\edgetoblobtransfer}{\ensuremath{upld(k)}}
\newcommand{\cloudfuncstart}{\ensuremath{start(m)}}
\newcommand{\cloudwarmstart}{\ensuremath{start_w(m)}}
\newcommand{\cloudcoldstart}{\ensuremath{start_c(m)}}
\newcommand{\cloudcomputetime}{\ensuremath{comp(k,m)}}
\newcommand{\cloudtoblobtransfer}{\ensuremath{store(k)}}
\newcommand{\edgecomputetime}{\ensuremath{comp(k)}}
\newcommand{\edgetoiothub}{\ensuremath{iotup(k)}}
\newcommand{\iothubtoblobtransfer}{\ensuremath{store(k)}}

\newcommand{\Cmax}{\ensuremath{\mathcal{C}_{max}}}

\newcommand{\surplus}{surplus}

\newcommand{\sep}[1]{}

\newcommand{\sedit}[1]{#1}
\newcommand{\del}[1]{}

\begin{document}

\title{Performance Optimization for Edge-Cloud Serverless Platforms via Dynamic Task Placement}
%\author{
%%Anirban Das$^1$, Shigeru Imai$^1$, Mike P. Wittie$^2$, Stacy Patterson$^1$ \\
%%\small {\em  $^1$Dept. Computer Science, Rensselaer Polytechnic Institute, Troy, New York 12180, USA}\\
%%\small {\em  $^2$Gianforte School of Computing, Montana State University, Bozeman, Montana 59715, USA} \\ [2mm]
%\small Submission Type: Research
%}

\author{\IEEEauthorblockN{Anirban Das, Shigeru Imai, Stacy Patterson}
\IEEEauthorblockA{
Dept. of Computer Science, Rensselaer Polytechnic Institute\\
dasa2@rpi.edu, shigeru.imai1@gmail.com, sep@cs.rpi.edu}
\and
\IEEEauthorblockN{Mike P. Wittie}
\IEEEauthorblockA{
Gianforte School of Computing, Montana State University \\
mike.wittie@montana.edu}
%\and
%\IEEEauthorblockN{3\textsuperscript{rd} Given Name Surname}
%\IEEEauthorblockA{
%%\textit{dept. name of organization (of Aff.)} \\
%\textit{Rensselaer Polytechnic Institute}\\
%Troy, USA \\
%email address or ORCID}
}

\maketitle

\begin{abstract}

We present a framework for performance optimization in serverless edge-cloud platforms using dynamic task placement. We focus on applications for smart edge devices, for example, smart cameras or speakers, that need to perform processing tasks on input data in real to near-real time. Our framework allows the user to specify cost and latency requirements for each application task, and for each input, it determines whether to execute the task on the edge device or in the cloud. Further, for cloud executions, the framework identifies the container resource configuration needed to satisfy the performance goals. We have evaluated our framework in simulation using measurements collected from serverless applications in AWS Lambda and AWS Greengrass. In addition, we have implemented a prototype of our framework that runs in these same platforms. 
%A key contribution of our work is the development of accurate, data-driven predictive models for edge and cloud serverless application pipelines.
In experiments with our prototype, our models can predict average end-to-end latency with less than 6\% error, and we obtain almost three orders of magnitude reduction in end-to-end latency compared to edge-only execution.
\end{abstract}

%%%%%%%%%%%%%%%%%%%%%%%%%%%%%%%%%%%%%%%%%%%%%%%%%%%%%%%%%%%%%%%%%%%%%%%%%%%%%%%%
\section{Introduction} 
\label{sec.introduction}
%
%Standard format:
%Brief introduction
%Some references 
%Challenges
%	1st challenge
%	1st approach
%	2nd challenge
%	2nd approach
%Why are we modeling this	
%Cold Start warm start mention.
%Contributions of this paper

In the past several years, there has been increased usage of intelligent applications on end-user devices, including voice-activated virtual assistants, smart-home security cameras, augmented reality in mobile phones, and so on. In many of these applications, the bulk of data processing is performed in a cloud data center.
The increase in the number of such applications, as well as in the number devices, has  led to a  surge in the amount of data generated at the periphery of communication networks~\cite{gartner21billion}.  This surge presents challenges to the cloud-based computational approach, as applications must compete for limited network resources~\cite{edgevision}. The situation is further complicated by the fact that many intelligent applications are latency sensitive, and thus require near real-time access to computational resources.

Edge computing proposes to address these challenges by leveraging the computational power in close proximity to data sources, sometimes in the end user devices themselves.
These \emph{edge devices} can perform data processing like filtering, aggregation, or inference. The results, typically much smaller in size, may then be forwarded to the cloud for further processing and decision-making. 
This approach reduces both data processing latency and bandwidth usage~\cite{satyanarayanan2017emergence, edgevision}.

To achieve this latency reduction requires that edge devices execute compute tasks in near-real time. 
Depending on the type of application or workload, this may not be possible on resource-constrained edge devices. In such cases, it may be necessary to offload  the computation to a higher-resourced compute node in the cloud.  The problem is then to determine which tasks should be executed on the edge device and which should be executed in the cloud, so as to meet developer-specified performance criteria such as latency or cost.
%~\cite{mec_offloading_survey,maui, kang2017neurosurgeon, shi2014cosmos}.

We study this task placement problem in the context of  serverless computing, more specifically, the Function-as-a-Service paradigm, an increasingly popular model for both cloud and edge platforms~\cite{serverlessicdcs,baldini2017serverless, edgebench}. 
In this paradigm, the developer writes stateless functions that can be triggered by various events.
Each function executes in its own container.
The developer specifies the container resource allocations, and the containers in the cloud are orchestrated and provisioned by the cloud provider.
The function performance depends on the input and application characteristics, the network transfer from the edge device to the cloud, the container resources, and the time to store the function results. 
 Several cloud platforms also provide frameworks for function execution on edge devices~\cite{greengrass_dev_guide, azure_dev_guide}.
 
We propose a framework that dynamically determines where to execute serverless functions so as to optimize developer-specified performance criteria. 
 Our framework targets intelligent applications that consist of a single serverless function that executes a data processing task on an input, for example, image recognition on a single frame (image) from a camera. %The results of the function are saved to cloud storage for further analysis. 
 Our framework processes a sequence of inputs, and for each input, it dispatches the task to a function in the appropriate container, from among the edge container and a set of containers in the cloud with various resource allocations.
 
Our framework addresses two optimization problems: (1) minimize  latency subject to a cost constraint and (2) minimize cost subject to a latency constraint, where the latency is measured in terms of the time from the ingestion of the input to the storage of the results in the cloud.
To perform this  dynamic task placement, our framework predicts the application latency and cost for the various container configurations and input characteristics. Thus, a key contribution of our work is the development of accurate, data-driven performance models for each component in the  edge and cloud execution pipelines. These models encompass network transfer time, container startup time, function execution duration, and storage latency.

The specific contributions of this work are as follows: (1) we present application-specific performance models for serverless applications in an edge-cloud computing platform - these models are trained and evaluated using data collected from applications running in the Amazon Web Services~(AWS) serverless environment; (2)  we propose a dynamic task placement framework, using these models, that optimizes for developer-specified metrics such as latency or cost; (3) we provide extensive evaluations of our framework using real-world data from applications running in AWS; and (4) we present a prototype implementation of our framework and its evaluation results.
In our experiments in AWS, our prototype predicts end-to-end latency with less than 6\% error.
In this study, we use AWS Greengrass for edge computing and AWS Lambda for the the cloud platform, but our framework can be  generalized to other serverless platforms.
% like Azure Functions~\cite{azure_func} or Google Cloud Functions~\cite{google_func}.

The rest of the paper is organized as follows. 
Sec.~\ref{sec.sysarchitecture} provides details about the system architecture and our benchmark serverless applications.
In Sec.~\ref{sec.framework_overview}, we describe our framework architecture and objectives.
Sec.~\ref{sec.perfmodel} presents the performance models and gives details of their training and evaluation.
In Sec.~\ref{sec.framework}, we give the details our framework implementation, and
in Sec.~\ref{sec.setup_results}, we present evaluation results  from both data-driven realistic simulations and a live prototype.
In Sec.~\ref{sec.relatedwork} we discuss related work, and we conclude in Sec.~\ref{sec.conclusion}.

%%%%%%%%%%%%%%%%%%%%%%%%%%%%%%%%%%%%%%%%%%%%%%%%%%%%%%%%%%%%%%%%%%%%%%%%%%%%%%%%
\section{Architecture and Applications} \label{sec.sysarchitecture}

A serverless edge-cloud architecture consists of an edge device with access to input data. The device is connected via a network to a cloud data center that has both compute and storage capability. Industry frameworks support two approaches or \emph{pipelines} to execute an application in this architecture, a \emph{cloud pipeline}, where the data processing task executes in the cloud data center and an \emph{edge pipeline}, where the task executes on the edge device itself. 
We briefly describe each pipeline below, as well as the factors that impact its end-to-end latency and execution cost.
We then describe the benchmark applications that we use to validate our models and framework.

\begin{figure}[htbp]
\centering
\begin{subfigure}[c]{.90\linewidth}
      \centering
      \includegraphics[width=1\linewidth]{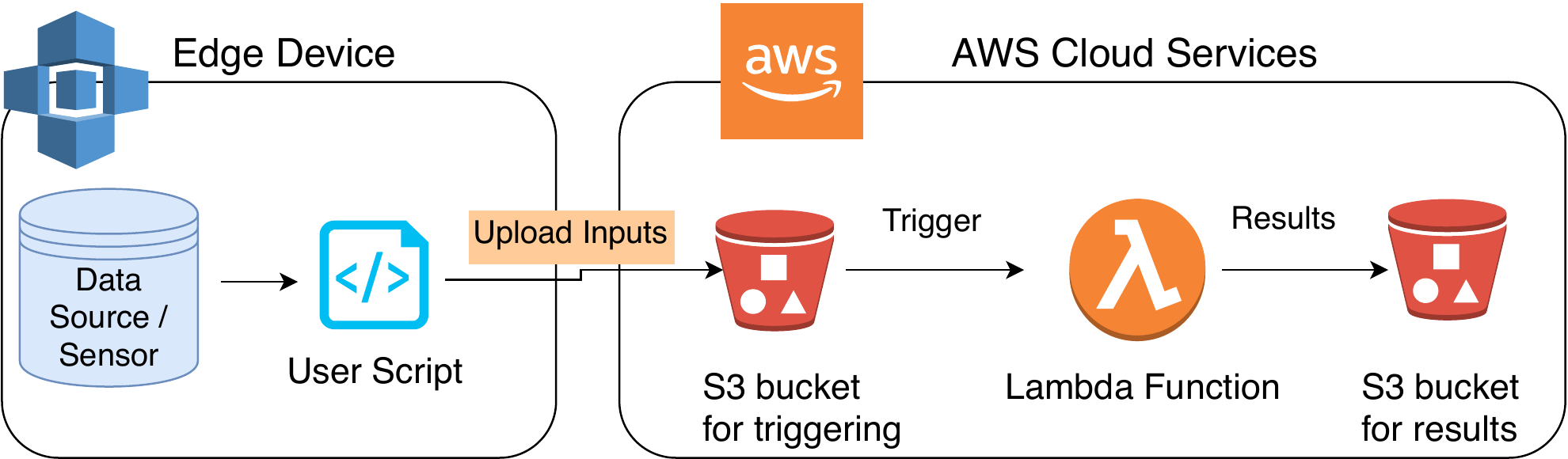}
      \caption{Cloud pipeline in AWS Lambda.}      \label{fig.lambda_architecture}
     \end{subfigure}
\begin{subfigure}[b]{.9\linewidth}
      \centering
      \includegraphics[width=1\linewidth]{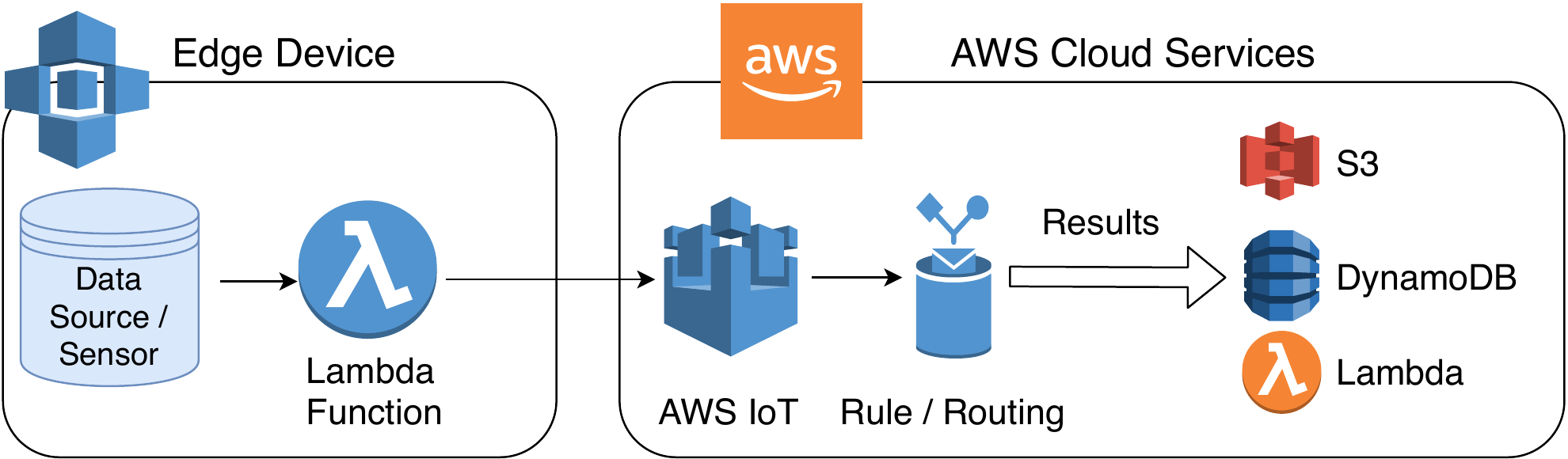}
      \caption{Edge pipeline in AWS Greengrass. } \label{fig.Greengrass_architecture}
\end{subfigure}
\caption{AWS serverless application pipelines.}
\end{figure}
\vspace{-.3cm}
\subsection{Serverless Computing Pipelines} \label{sub.pipelines}

\subsubsection{Cloud Pipeline} \label{subsub.cloud_pipeline}

% summary of pipeline
In the cloud pipeline, the edge device uploads the input data to a cloud storage service. This upload triggers the execution of the stateless function that performs the data processing task.
The function writes the result of this task to cloud storage. An example of this pipeline for AWS Lambda is shown in Fig.~\ref{fig.lambda_architecture}. Here, the edge device uploads input data to an AWS S3 bucket, triggering an AWS lambda function that stores the result in another S3 bucket.

% Fig.~\ref{fig.lambda_architecture} shows a typical cloud application architecture using AWS Lambda functions. A lambda function is a piece of event triggered, stateless user code running inside a container \cite{aws_lambda_internals}. Generally, input data is uploaded to an AWS S3 bucket, triggering the lambda function that performs computation on the data and stores result in pre-configured data sink(s). Lambda functions can be pre-configured with different memory settings during initial deployment and can be arranged to be triggered from specific S3 buckets. AWS assigns CPU power to the lambda functions proportional to their memory configuration. At runtime, based on computational requirements, data can be uploaded to specific S3 buckets to trigger different lambda functions. 
%\note{How are these different configurations selected at runtime?}

% resource allocation options
In commercial serverless platforms, each function instance runs in its own container. 
Further, some platforms allow developers to configure the container resources, e.g., memory and CPU, during application deployment.
This resource configuration, in turn, impacts the function performance when executed in that container. 
For the purposes of this work, we adopt the container configuration options offered by AWS. The developer specifies the memory allocation for the container, ranging from 256~MB to 3008~MB.
%The developer specifies the memory allocation for the container, ranging from X to X MB.  
AWS assigns CPU power to the container proportionally to the allocated memory.

%\container creation and parallel execution
To scale serverless applications, serverless platforms create containers as needed.
 When a function is triggered, if a container is available (not currently executing a function), the function executes in the existing container. Otherwise, a new container is created for that function execution.  If a container is idle for some extended period, it is destroyed.  If a new container needs to be created before executing a function, this is called a \emph{cold start}. A cold start imposes
 a non-trivial time penalty to initialize the container and load any libraries before the function execution begins. If a function is assigned to an existing container, this is a \emph{warm start}, and the startup time can be  up to an order of magnitude lower than a cold start. 
% \sep{Anirban, would it be correct to say 'up to an order of magnitude less'? Or if not, is there a percentage less for warm start that we can state here?}
 %\sq{Note, we may not need to italicize warm and cold start if they are already italicized in the introduction}

%A factor influencing end to end latency of serverless functions is the concept of `cold start'~\cite{aws_lambda_internals}\cite{lloydlatencyserverless}\cite{manner2018cold}. Cloud providers recycles or kills the containers for lambda functions if they are idle for an extended period of time. The next time they are invoked, there exists a non-trivial time penalty needed to first initialize a container for the function and then the libraries etc. This is called a `cold start' latency.  Also, if all existing containers for a lambda function are busy executing workload, invoking the lambda function will result in cold start due to creation of a new container to scale up. However, when a lambda functions is in a `warm' state, i.e. a container already exists, it takes much less time to invoke that function. In order to perfect our end to end cloud latency prediction models Sec.~\ref{sec.perfmodel}, we need to include the effects of cold start and warm start by modeling the cold start performance as well.

% definition of end-to-end latency
\paragraph{Pipeline performance}We measure the performance of the cloud pipeline in terms of the \emph{end-to-end latency} for processing a single input.  
This measurement begins when the input is ingested by the application on the edge device and ends when the result is saved in cloud storage (the second S3 bucket in Fig.~\ref{fig.lambda_architecture}).
The latency, for an input $k$ and memory configuration $m$, consists of the following components:
\begin{itemize}[noitemsep,topsep=0pt]
\item \edgetoblobtransfer : The time to transfer the input from the edge device to cloud storage. This consists of the network transfer time and any write overhead to the first S3 bucket.
% S3 bucket write time overhead.
%as well as any time overhead associated with storing the input in the cloud. 
\item \cloudfuncstart : The time to start the cloud container for function execution. The time depends on the container memory $m$, but not the input $k$ and varies for a cold start or warm start. 
%\sep{make sure to give cold and warm start notation later.}
%denoted by \cloudcoldstart \;or a warm start \cloudwarmstart.
\item \cloudcomputetime : The compute time in the function.
\item \cloudtoblobtransfer : The time to save the output to cloud storage. 
\end{itemize}

The end-to-end latency is therefore:
\begin{equation}
T_c(k)= \edgetoblobtransfer + \cloudfuncstart + \cloudcomputetime+ \cloudtoblobtransfer. \label{eqn.cloud-latency-model}
\end{equation}

\paragraph{Pipeline cost} 
In serverless cloud platforms, the cost is typically based on the function execution time. In this work, we use the AWS pricing model.
AWS bills users for the duration for which code executes on AWS systems, rounded up to the nearest 100\,ms. The price is proportional to the amount of memory allocated to the container at \$$1.667 \times 10^{-06}$ per GB-s execution~\cite{aws_lambda_pricing}. Further, there is a fixed charge of \$0.20 per 1\,M lambda function requests. 
To determine function execution cost from execution time \cloudcomputetime, we round the execution time to the nearest ms and then apply the AWS pricing model.
We limit our study to the function execution cost, as this is the most challenging to model and predict. 
% \sep{maybe remove the rest of this paragraph, and put the first sentence at the end of the previous paragraph.}
% \das{This can be elaborated if we have space. }
%While there are charges to use S3 for the input data, 
%this intermediate cloud storage may not be long-lived and could be implemented using a variety of data forwarding options within AWS. The cost of the results storage is S3 is the same in both pipelines.
%We also do not include the cost to transfer data from the edge device to the cloud.
%In many networks, there are no upload caps, and so the upload cost may be amortized to zero over time. 

\subsubsection{Edge Pipeline} \label{subsub.edge_pipeline}
One can also use an edge pipeline, where data processing is performed within a function  on the edge device itself. 
Upon function completion, only the results, usually much smaller in size, are sent to the cloud for storage.
% (second S3 bucket) \sep{I don't understand what is meant by second S3 bucket here? I suggest removing it. Cloud storage is more general than S3, as indicated in the next paragraph}. 
Similar to the cloud platform, the developer has a facility to constrain the memory limit of the lambda function, but here the upper limit is dictated by the available resource in the edge device. 
%and is fixed for the lifetime of an application. 
Thus, we assume a single memory configuration for the edge device container.

In Fig.~\ref{fig.Greengrass_architecture}, we show an example of an edge pipeline using the AWS Greengrass edge computing framework.
Similar pipelines can be created in other edge computing platforms, e.g. Azure IoT Edge.
In Greengrass, a lambda function executes inside the Greengrass run-time on an edge device, and the function results are subsequently sent to the AWS IoT Core service in the cloud. The IoT Core service forwards the results to a developer-specified endpoint, for example, S3 or DynamoDB. 
%\sep{I want to move these next few sentences to a place where we describe the experiments/data collection. Here, we are describe abstract pipelines, and these details detract from the abstraction.}
%For this study, we use an S3 bucket as the endpoint for the edge pipeline, just as in the cloud pipeline, with one exception. Currently, AWS Greengrass, only supports upload of \texttt{json} encoded results; if the result of the function is a binary file, for example, an image, we can directly upload the file to the S3 bucket, bypassing the IoT Core service.

Greengrass offers two execution models for lambda functions, a stateless function and a `long-lived' function~\cite{greengrass_dev_guide}. The stateless model is similar to the model used in the AWS Lambda platform, in that multiple functions may execute on an edge device in parallel. In the long-lived model, the function runs continuously on the edge device; it can write to and read from device storage and this storage persists for the lifetime of the function. We use the long-lived function model for several reasons. First, we consider resource-constrained edge devices that may not have the power to execute functions in parallel, depending on the application.
%Our edge device is resource constrained \- running multiple lambda functions may be very resource intense, it is difficult to determine how many parallel invocations can be handled by the device depending on the application. 
Second, co-location of multiple functions contending for limited hardware resources may cause unpredictable behavior.
This unpredictability limits the ability to optimize task placement.
%which makes it difficult to determine optimal task placement.

%However, we plan to tackle this avenue in future study.

%In our implementation, these `local' lambda functions ingest the workload from the input data source and perform necessary processing. 
%\note{What do you mean by `local' here. Is a local lambda function a specific thing?}.
 %It's also possible to directly connect to other AWS services from the edge device for e.g. to upload resized images to S3 bucket in Sec.~\ref{subsec.applications}. 

\paragraph{Pipeline performance}
We measure the performance of the edge pipeline in terms of the end-to-end latency for processing a single input. 
%The latency measurement begins when the input is ingested in the edge device and ends when the result is available in the cloud storage. 
The latency for an input $k$ consists of the following components: 
    \begin{itemize}[noitemsep,topsep=0pt]
    	\item \edgecomputetime: The compute time on the edge device.
		\item \edgetoiothub: The time to send the results from the edge device to the cloud IoT Core, including the network transfer time and the framework-induced overhead. 
		\item \iothubtoblobtransfer: The delay between when the results are received in the cloud IoT Core service and when they are available in the cloud storage.
  	\end{itemize}
The end-to-end latency for the edge pipeline is:
\begin{align}
T_e(k) = \edgecomputetime + \edgetoiothub + \iothubtoblobtransfer. \label{eqn.edge-latency-model}
\end{align}
%We estimate the end-to-end latency of an edge execution by modeling each of these components separately, as described in Sec.~\ref{subsec.edge-performance-model}

\paragraph{Pipeline cost}
For the edge pipeline, we again consider the AWS pricing model. Lambda function execution inside AWS Greengrass is free, but there exists a fixed yearly device registration fee of \$1.49 - \$2.05 based on the region. The cost is fixed per active edge device per month and is independent of the number of function executions.
Thus, we consider the amortized function execution cost at the edge to be zero.
Our cost analysis excludes storage and network costs, as processed data sent by edge devices is small and costs are easy to predict with respect to request volume.
% We also do not include the storage cost and internet provider specific-network costs, for reasons similar to those given for the cloud pipeline.

%--------------------------------------------------------
\subsection{Applications} \label{subsec.applications}
We implement three representative applications, motivated by real-world use cases.  
All are implemented in Python.

\textbf{Image Resizing (IR):} An image file is taken as input, and the function reduces the dimensions of the image to a 128$\times$128 pixel thumbnail and sends the thumbnail to the cloud.  This reduction can be done to save bandwidth or storage cost, or to regularize the image for use in a deep learning application. 
This application mimics a scenario where a traffic camera takes a stream of pictures to identify traffic congestion.
For the input workload, we use a set of images from the Image of Groups~\cite{gallagher_cvpr_09_groups} database. 
%\sep{Is this 1400 for model training and validation and 600 for simulations?} \das{Yes}

\textbf{Face Detection (FD):} Here, the application, given an input image, finds the number of faces present. The application mimics a smart camera  that detects faces in a captured frame, for either security purposes, for example.  For face detection, we use the \texttt{dlib}~\cite{dlib09} library, and for simplicity, we store only the number of faces detected in the frame. 
We use images from the Images of Groups database for our input workload. 
%\sep{Do we use the same number of images for this workload as for IR?} \das{Yes. Same.}

\textbf{Speech-To-Text (STT):}  An audio file containing speech is provided as input, and the application transcribes the speech into text. This emulates the functionality of a smart speaker where the user issues commands that are translated to text and then used as input in a search or activity.  
For the transcription, we use CMU's \texttt{pocketsphinx}~\cite{pocketsphinx} library.  
For our input workload, we use audio files from the Tatoeba Corpus \cite{tatoeba}.  
%\sep{How many?} \das{3400 for model training validation and 600 for experiments.}

\sedit{We note that different applications may have different input rates from the data source. While input commands to a smart speaker may be sparse, a traffic camera may produce a fixed number of images per minute. To simulate this behavior, the applications ingest input files from a local directory on the edge device at a fixed rate.
For IR and FD, we implement a faster input rate of  four files or images per second, and for STT, we use slower rate of one audio file every ten seconds.}

%%%%%%%%%%%%%%%%%%%%%%%%%%%%%%%%%%%%%%%%%%%%%%%%%%%%%%%%%%%%%%%%%%%%%%%%%%%%%%%%

\begin{figure}[htbp]
      \centering
      \includegraphics[scale=.40]{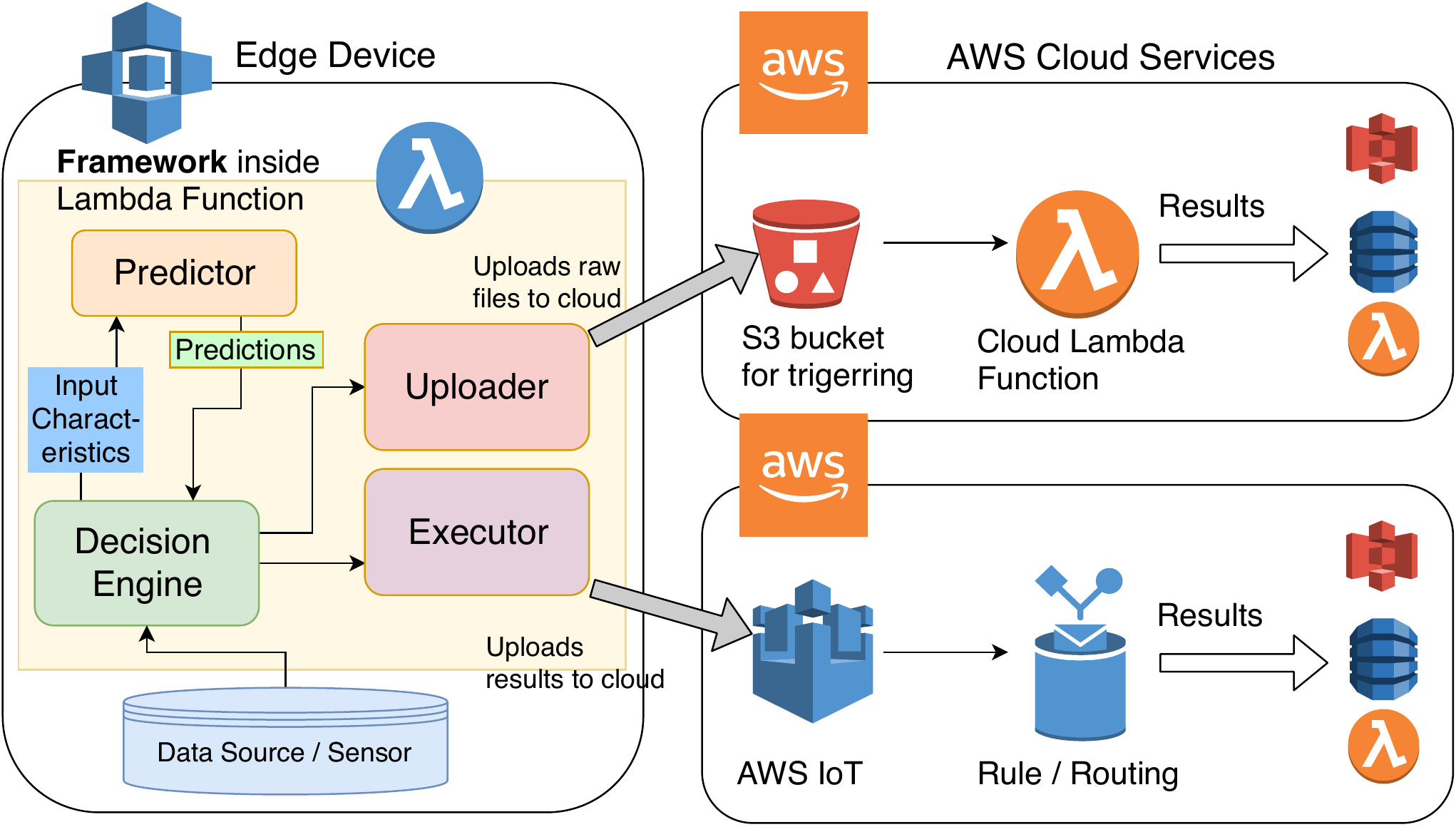}
      \caption{Dynamic task placement framework.}
      \label{fig.aws_edge_cloud_architecture}
      \squeezeuppicture
\end{figure}
\section{Framework} \label{sec.framework_overview} 
For applications such as those in Sec.~\ref{subsec.applications}, it is necessary to execute tasks in container configurations with appropriate resources to meet performance criteria. Hence, we introduce a framework that dynamically selects the cloud or the edge pipeline
to execute the application tasks (functions) based on the workload characteristics and latency or cost requirements.

\subsection{Framework Architecture}
Fig.~\ref{fig.aws_edge_cloud_architecture} depicts our framework architecture.
We consider a single edge device that ingests an input workload.
% according to some preset rate. \sep{Is a preset rate necessary for the framework to work? Could it be an arbitrary, changing rate?} \das{Theoritically the rate can be changing because there is nothing in the algorithms which explicitely depends on the rate. The results may vary. But here we did not study for varying incoming inputs.}
The edge device runs a single lambda function. 
%We denote this configuration by $\lambda_{edge}$. 
There  are also $N$ lambda functions configured in the cloud with $N$ distinct memory configurations.
Let $\Phi = \{\lambda_{m_1}, \lambda_{m_2}, \cdots \lambda_{m_N} \}$ be the set of $N$ cloud container configuration options,
where container type $\lambda_m$ is configured with $m$ MB of memory.

%We denote these cloud container configurations by  $\lambda_m$, where $m$ is the memory configuration in MB.
%Each cloud lambda function has its own S3 buckets, which, when a file is uploaded to it, triggers the execution of that lambda function.  
%%We let $\pmb{K}$ be the total number of inputs / jobs per application \sep{Why do we define $K$?}. 
%Throughout this study we use $\lambda_m$ to denote a lambda function memory configuration of $m$~MB in the cloud and $\lambda_{edge}$ to denote the single lambda function running on the edge device. Finally, let $\Phi = \{\lambda_{m_1}, \lambda_{m_2}, \cdots \lambda_{m_N}, \lambda_{edge}\}$ be the set of N+1 possible sites for task placement i.e. execution sites.

The framework functionality resides in the lambda function on the edge. It ingests input workload from the Data Source. 
The data then flows to the Decision Engine. The Decision Engine first calls the Predictor; given an input, Predictor predicts the end-to-end latencies and costs for executing the application  for each configuration in $\Phi$, as well as for executing the data processing task in the edge device. \sedit{Due to the differences in cold start and warm start times, the Predictor must predict both whether an available container exists for a given cloud configuration, and what the performance of the function will be in that configuration.}

The Decision Engine, then, based on the objective (Sec.~\ref{subsec.problem_formulation}) and the predicted latencies and costs, either places the task at the edge  or selects a $\lambda_m$ in cloud to offload the task. If the edge is selected,
 the job is sent to the Executor, which contains a FIFO task queue. It executes tasks one at a time and sends the results to the cloud.
We note that the Executor in Fig.~\ref{fig.aws_edge_cloud_architecture} corresponds to the lambda function in Fig.~\ref{fig.Greengrass_architecture}. With some abuse of terminology, we refer to the Executor as $\lambda_{edge}$.
If a cloud configuration is selected, the Uploader uploads the input file to the corresponding S3 bucket in the cloud. The rest of the standard cloud pipeline then continues as in Fig.~\ref{fig.lambda_architecture}. 
We configure the edge function execution to be non-blocking, so that the Decision Engine processes each input without waiting for the completion of the previous task. 
\sep{I commented out the next two sentences because I think this is adequately explained in Section 5.}
%If the Executor is busy and there exists $ \lambda_m \in \Phi$ that satisfies the performance requirements and the execution time, including wait time in queue at edge is larger than the execution time at $\lambda_m$, the Decision Engine will choose that $\lambda_m$. Otherwise, the input will be added to the Executor queue.

\subsection{Optimization Objectives}\label{subsec.problem_formulation}
Our framework provides two options for task placement policy.
We detail these policies below.

\paragraph{Cost minimization subject to deadline constraints}
%An offline optimal solution to this problem of reducing the total cost subject to end-to-end latency deadline per job requires the knowledge of the entire input sequence and does not fit well in a streaming scenario. If all information is available beforehand, a solution can be obtained solving a binary integer linear program, but it will have exponential complexity with the number of decision variables i.e. the choices of execution sites.
The objective is to minimize the execution cost, subject to developer-specified end-to-end latency deadline $\delta$ per task. 
%We greedily minimize predicted cost per job subject to predicted end-to-end latency under the time constraint to minimize the total cost of execution. 
%Let $T(k)$ denote the end-to-end latency for input $k$, and let  $\mathcal{C}(k)$ be the incurred cost.
% \sep{maybe remove this formal problem definition. I think it is obvious from the text.}%The problem is then:
%\begin{align}
%\label{subsub.Problem1}
%\begin{split}
%		\text{minimize}~& C(k)  \\
%		 \text{subject to}~& T(k) \leq \delta\, .
%\end{split}
%\end{align} 
The framework achieves this goal by, for each task $k$, selecting the least expensive configuration that satisfies the deadline. 
This formulation targets applications with strict latency requirements, for example, an application that uses a smart camera to monitor a secure site and send alerts on detection of occupants for further analysis.
Note that for this problem to have a feasible solution, there must be at least one configuration that can process any input within the given deadline.

%We use the models from Sec.~\ref{sec.perfmodel} to select the site that results in the minimum.
\paragraph{Latency minimization subject to cost constraints}
Here, the objective is to minimize the end-to-end latency while keeping the cost of each task execution under some budget $\Cmax$.
For example, a store may use a smart camera to recognize customers and text them coupons.
 In this case, the application providers may have a strict budget, and while low latency is desirable, it is not necessary.
If we consider only a single input $k$, the goal is then to select a configuration that solves the problem:
\begin{align} \label{subsub.Problem2}
\begin{split}
		\text{minimize}~~& T(k) \\
		\text{subject to}~~& C(k) \leq \Cmax \, ,
\end{split}
\end{align}  
where $T(k)$ and $C(k)$ are the latency and cost for task $k$, respectively.

It is possible that single function execution may not use up the entire budget $\Cmax$, i.e., a sequence of tasks may leave budget surplus  that could have been used to reduce latency. So, instead, we consider the constraint that for any sequence of $K$ tasks, we have $\sum_{k=1}^K C(k) \leq K \Cmax$.
We implement this constraint using the following optimization problem:
\begin{align} 	\label{opt:optim_time_cost_surplus}
\begin{split}
\text{minimize}~~& T(k) \\
\text{subject to}~~& C(k) \leq \Cmax + \alpha  \times \surplus(k)
\end{split}
\end{align}
where $surplus(k)$ is the accumulated sum of unused budget, $\surplus(k) = \sum_{i=1}^{k-1} (\Cmax - C(i))$, and
$\alpha \in [0,1]$ is a scaling factor that determines how much of the surplus can be used for task $k$. 
Since the cost of executing the task at the edge is zero, it is always possible to find a task placement that satisfies the cost constraint. Thus, the surplus is never negative.

\section{Performance Models} \label{sec.perfmodel}
To solve the task placement problems described in Sec.~\ref{subsec.problem_formulation} requires accurate methods for predicting the end-to-end latency $T(k)$ 
for any input $k$. From this predicted latency, we can also compute the predicted cost $C(k)$. 
To predict the latency, we create data-driven performance models for each latency component described in Sec.~\ref{sub.pipelines}.
Due to the heterogeneity of application characteristics, we create \emph{application-specific} performance models, trained using sample input data.
In this section, we describe our performance models and model training, and we give results on the model accuracy.

\subsection{Cloud Performance Model} \label{subsec.cloud-performance-model}
We segregate the end-to-end latency $T_c(k)$ for job $k$ configuration $\lambda_m$ into four parts, as described in Eqn. (\ref{eqn.cloud-latency-model}).

%and use either mean or regression models for each and them combine them together for the end to end latency. We observe Upload time ($\edgetoblobtransfer$) from edge device to cloud storage depends on the input data size. We fit a linear regression over measurements of input upload time. The Lambda startup time  varies depending on whether it is a cold start, denoted by $\cloudcoldstart$, or a warm start, denoted by $\cloudwarmstart$.
%Based on training data,  we observe that each of these startup times follows a normal distribution and estimate each of these times to be the mean over the training data.
%Compute time ($\cloudcomputetime$) is observed as to be a non-linear function of $size(k)$ and the container memory configuration $m$. After experimenting with several  regression methods,  we selected Gradient Boosted Regression Trees~\cite{friedman2002stochastic} to model $\cloudcomputetime$. Finally, for Storage time ($\cloudtoblobtransfer$), we note the sizes of the function results are both small and very similar across applications. Further, because AWS S3 quantizes the file availability timestamp to seconds, 
%we are only able to measure the storage time with coarse granularity (also observed in Sec~\ref{subsec.edge-performance-model}). 
%Thus, our training data exhibits no correlation between the input and the storage time.
%We, therefore, model the storage time as a quantized normal random variable, and we model $\cloudtoblobtransfer$ by taking the mean over the training set.

\begin{itemize}
\item{\emph{Upload time:}} We model the upload time as a linear function of the input data size:
\[
\edgetoblobtransfer = \theta_1 + \theta_2\times size(k),
\]
%where 
$\theta_1$ and $\theta_2$ are determined via regression over training data.

\item{\emph{Lambda startup time:}} The startup time varies depending on whether it is a cold start $\cloudcoldstart$, or a warm start $\cloudwarmstart$.
Based on training data,  we observe that each of these startup times follows a normal distribution, which we model by taking the mean of the training data.

\item{\emph{Compute time:}} We observe that the compute time is a non-linear function of $size(k)$ and the container memory configuration $m$. After experimenting with several  regression methods,  we identified Gradient Boosted Regression Trees~\cite{friedman2002stochastic} to be the most accurate.

\item{\emph{Storage time:}} The sizes of the function outputs are both small and very similar across applications. Further, because AWS S3 quantizes the file availability timestamp to seconds, 
we are only able to measure the storage time with coarse granularity.
% (also observed in Sec~\ref{subsec.edge-performance-model}). 
Thus, our training data exhibits no correlation between the input and the storage time.
We, therefore, model the storage time as a quantized normal random variable, and we model $\cloudtoblobtransfer$ by taking the mean over the training set.

\item{\emph{Container idle time:}} To predict whether invoking a function will cause a warm start or a cold start, we also need to model the container idle time, i.e., how long containers stay warm in AWS infrastructure before their resources are reclaimed due to inactivity. We observe that this idle time is independent of any input or application characteristics and model this by a single value $T_{idl}$. We perform experiments, similar to the approach taken in \cite{wang2018peeking}, and use a binary search to find $T_{idl}$. Our findings corroborate with this previously measured value of $T_{idl} \approx 27 $ minutes.
\end{itemize}

\subsection{Edge Performance Model} \label{subsec.edge-performance-model}
For edge pipelines, we also model the components of Eqn.~\eqref{eqn.edge-latency-model}  separately. 

\begin{itemize}
\item{\emph{Compute time:}} We model the compute time as a linear function of the input file size:
\[
\edgecomputetime = \phi_0 + \phi_1 \times size(k). 
\]
We determine the parameters $\phi_0$ and $\phi_1$ using regression over the training data.

\item{\emph{IoT Core upload time:}} As previously mentioned, the size of function results are small and similar across inputs and applications.
Thus, we attribute any variability in recorded times to framework and network overhead. We model this upload time as a normal random variable, and we estimate \edgetoiothub\ by the mean over the time measurements of the training data.

\item{\emph{Storage time:}} We adopt a similar approach to the storage time for cloud pipeline, and estimate \iothubtoblobtransfer\ using the mean over the measurements from the training data. 
\end{itemize}

%We observe Compute time (\edgecomputetime) on the edge device, depends linearly on the file size and subsequently model using linear regression over the training data, as described in Sec.~\ref{sec.model_exps}. For, IoT Core upload time, as previously mentioned, since the size of function results are small and similar across inputs and applications, we attribute variability in recorded times to framework and network overhead. We model this upload time as a normal random variable, and we estimate \edgetoiothub\ by the mean over the time measurements of the training data.
%Finally, for Storage time (\iothubtoblobtransfer) at the edge, we adopt a similar approach to the storage time for cloud pipeline, and estimate  using the mean over the measurements from the training data. 

%%%%%%%%%%%%%%%%%%%%%%%%%%%%%%%%%%%%%%%%%%%%%%%%%%%%%%%%%%%%%%%%%%%%%%%%%%%%%%%%
%\subsection{Model Accuracy Evaluation and Results}\label{sec.model_exps}
\subsection{Model Training and Evaluation}\label{sec.model_exps}
%Here, first we describe our experimental setup, then 
%Here, we describe how we collect training data, i.e., latency measurements on sample inputs, and how we use this data to train the models. We also present evaluation results demonstrating the accuracy of our models.

To generate the measurements for training and evaluation, we execute the applications described in Sec.~\ref{subsec.applications} using the pipelines shown in Fig.~\ref{fig.lambda_architecture} and Fig.~\ref{fig.Greengrass_architecture}.
For all experiments, for the cloud pipelines, we use 19 AWS Lambda function memory configurations between 640~MB and 3008~MB. % with 128~MB intervals except 2816~MB. 
For the edge device, we use a Raspberry Pi 3B running Greengrass core version 1.7.0.  The edge device container is provisioned with 512\;MB RAM and set to run indefinitely. 
 The image and audio file directory and the directory for storing metrics are mounted into the Greengrass execution environment as `Local Resources'. 
 The edge device is connected to the internet via a wireless router using the 2.4 GHz spectrum. A dedicated Stratum 1 NTP time server, TM2000A~\cite{tm2000A} 
is used to synchronize the time of the Raspberry Pi to AWS servers. 
% It is known that AWS also uses a highly accurate time synchronization service for its services.

%--------------------------------------------------------
\subsubsection{Cloud pipeline data collection} \label{cloudcollection.sec}
%--------------------------------------------------------

To train the components of the cloud latency model,  we first collect measurements of 
the  \edgetoblobtransfer, \cloudcoldstart,  \cloudwarmstart, \cloudcomputetime, and \cloudtoblobtransfer\  by running the pipeline using only warm starts.
%To collect measurements from the warm start execution for a given container configuration, 
We ensure there is an available container by first executing the function on a dummy input.
 We then upload each input file to the specified S3 bucket. This upload triggers the execution of the lambda function. We wait for a time interval in between uploads to ensure the previous function execution has completed. 
For each input, we measure each mentioned time component following a method similar to our previous work~\cite{edgebench}.
% For each input, to measure each component of the time mentioned above, we follow a method similar to our previous work~\cite{edgebench}.
\looseness-1

%For each input, we record  a timestamp $T_1$ when  the edge begins uploading the input file to S3. 
%When S3 finishes processing the upload event, AWS adds a timestamp $T_2$ and passes this timestamp to the function on its invocation. We save another  timestamp $T_3$ manually just after entering the lambda function.  At  the end of the function execution, we attach a timestamp $T_4$ to the results that are transmitted to S3. Timestamp $T_5$ is the AWS timestamp of the availability of the result in S3. 
%We measure \edgetoblobtransfer from the difference between $T_2$ and $T_1$. The startup time \cloudwarmstart\ is $(T_3 - T_2)$.
%The difference between $T_4$ and $T_3$ gives \cloudcomputetime. Finally, the difference between $T_5$ and $T_4$ is \cloudtoblobtransfer.

To measure the cold start time \cloudcoldstart \; for separate values of $m$, we use a method similar to that in \cite{lloyd2018serverless}. For each container configuration, we measure 100 cold start latencies. The cold start latency does not appear to be correlated to the container memory size for the three applications.

\begin{table}[htbp] 
\caption{Mean latencies (ms) used for training examples.} \label{means.tab}
\footnotesize
\centering
\begin{tabular}{|l|c|c|c|c|c|}
\hline
& \multicolumn{3}{|c|}{Cloud Pipeline}& \multicolumn{2}{|c|}{Edge Pipeline} \\
\hline
& Warm Start & Cold Start & Store & IoT Upload & Store \\
\hline 
IR  & 162  & 741 & 549 & n/a & 579 \\
\hline
FD & 163 &1500 & 584  & 25 & 583 \\
\hline
STT & 145 & 1404 & 533 & 27 & 579 \\
\hline
\end{tabular} 
\squeezeuppicture
\end{table}

%We obtain the values of parameters $\theta_{warm}(m)$ by taking mean over \cloudfuncstart \; for each lambda function memory $m$ and $\theta_{trnf}$ by taking mean over \cloudtoblobtransfer \; on the training set.
%: time to put the input file in S3: $T_c(k,i)^f$, warm start for the Lambda function: $T_c(k,i)^s$, and time to put results in S3: $T_c(k,i)^h$.
%The end-to-end warm start-latency for executing job $k$ in lambda function with memory $m$ is therefore, given according to Eqn.~\eqref{eqn.warm-start}.
%We used $\approx$ 4000 different audio files to get the metrics.

%
%We observe the Mean Absolute Percentage Error (MAPE) for predicting \edgetoblobtransfer \; ranges between 13.84\%- 16.78\% for the applications on test data. We use mean values for \cloudwarmstart / \cloudcoldstart\; corresponding to different memory configurations, however, overall we observe for \cloudwarmstart \;, IR has mean 162ms and standard deviation (std) 124ms, FD has mean 163ms and std 143ms and STT has mean 145ms and std 55ms. For \cloudcoldstart\;, we observe, overall coldstarts for IR has mean 741ms and std 264ms, for FD  with mean 1500ms and std 256ms, and the cold starts for STT has mean 1404ms and std 231ms.  Finally as an estimate of \cloudtoblobtransfer \; we observe the expected values lies around 549ms for IR, 584ms for FD and 533ms for STT.

\subsubsection{Edge pipeline data collection}
%\textbf{Model end to end time with Greengrass:} 
For each input workload, we measure the components in Eqn.~\eqref{eqn.edge-latency-model}: \edgecomputetime, \edgetoiothub, \iothubtoblobtransfer.
We set up the edge device running AWS Greengrass to ingest input files from a directory. Results are then sent to AWS IoT on the cloud, where a `Rule' redirects the results to an S3 bucket.  
\sedit{We note that for the IR application, we directly transmit the resized image to S3 due to Greengrass' limitations on data upload type.  Hence, we measure the storage time \iothubtoblobtransfer\ as the time between when a file is sent from the edge device to when it is available in S3,
and we do not include \edgetoiothub\ in the end-to-end latency.}

%From Eqn.~\eqref{eqn.edge-latency-model} we obtain the parameters $\phi^f,\phi^h$ by taking mean over \edgetoiothub \; and \iothubtoblobtransfer \; respectively on the training set. 
%Since local compute time, \edgecomputetime \; is dependent on a single feature i.e. the file property (e.g. file size or total number of pixels), we use simple Ridge Regression analysis from \texttt{scikit-learn}. 

\subsubsection{Model Training and Evaluation} \label{models.sec}

\sedit{For the IR and FD applications, we collect measurements for each configuration for 1400 input images. For the STT application, we collect measurements for 3400 input auto files for each configuration.}
We use the common 80:20 train:test split to train our models. For the regression techniques, we use a grid search for model tuning and chose the best performing estimator %configuration of hyper-parameters 
by using 3-fold cross-validation. 
In the FD and IR applications, we quantify \emph{size} by the total number of pixels in the image, as image manipulation depends on the matrix of the image pixel values. In STT, \emph{size} is measured in bytes.
 We use the gradient boosting regressor in the \texttt{scikit-learn}~%\cite{scikit-learn} 
 package for regression analysis of \cloudcomputetime \; and \edgetoblobtransfer \; respectively, for each application and ridge regression for modeling \edgecomputetime. 

Table~\ref{table.mape} shows the MAPE of the end-to-end latency over their respective test sets for the edge pipelines and cloud pipelines with warm start.
The error is less than 16\% for most of the applications. This suggests our models can predict the end-to-end latency relatively accurately. The one exception is the IR cloud pipeline, with MAPE of 25.3\%. 
%We observe that this higher error is due to the small magnitude 
%\sep{put mean here, if we remove details above}
% and large variance  
%\sep{put variance here if we remove the details above}
% of the compute time of IR pipeline. 

\begin{figure}%[tbp]
\begin{minipage}{\columnwidth}
    \centering
      \begin{subfigure}{0.45\linewidth}     
      \centering
      \includegraphics[width=\linewidth]{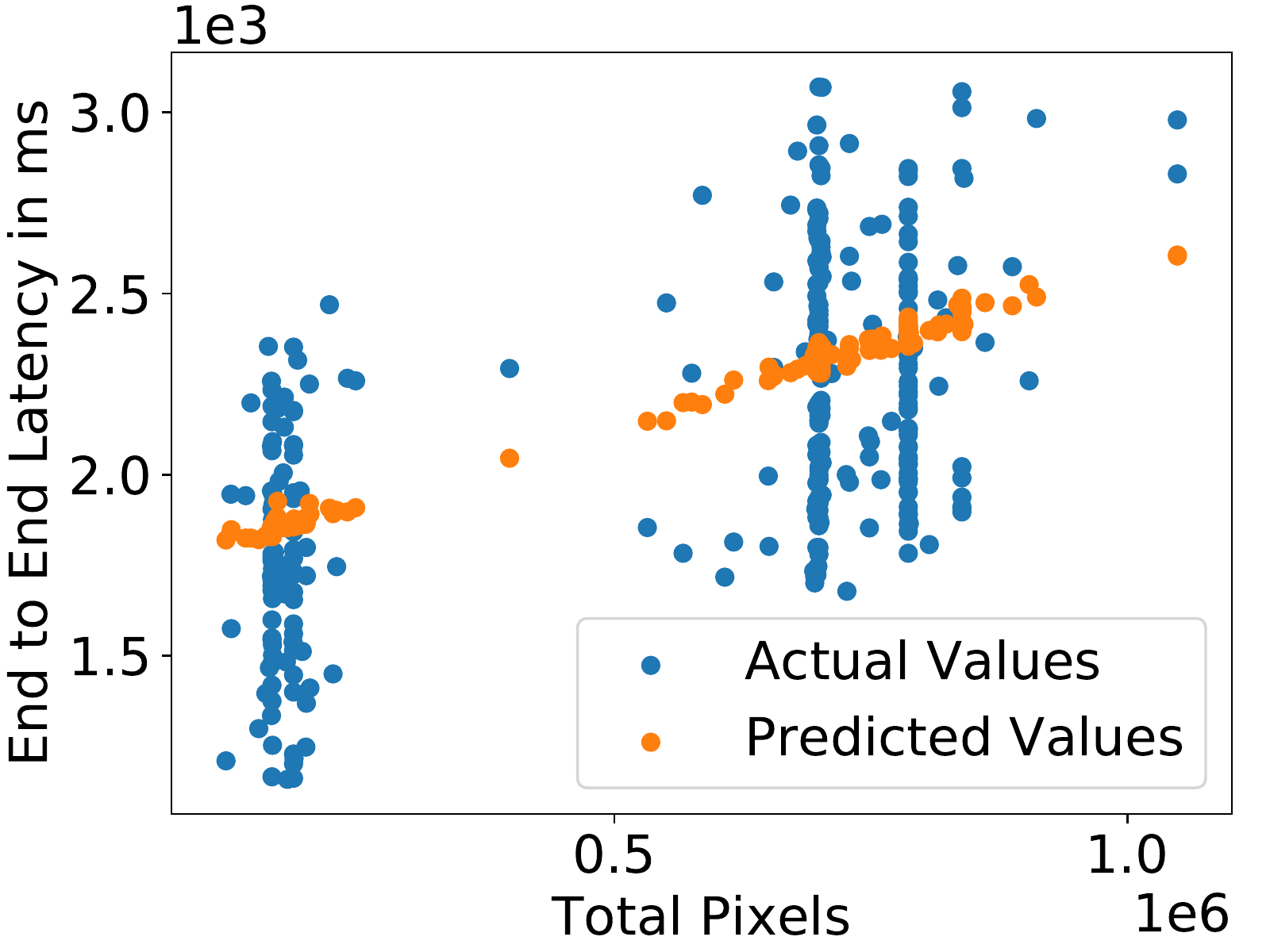}
      \caption{FD.}
      \label{fig.cloud_test_facedetect}
   \end{subfigure}%
   \begin{subfigure}{0.45\linewidth}
      \centering
      \includegraphics[width=\linewidth]{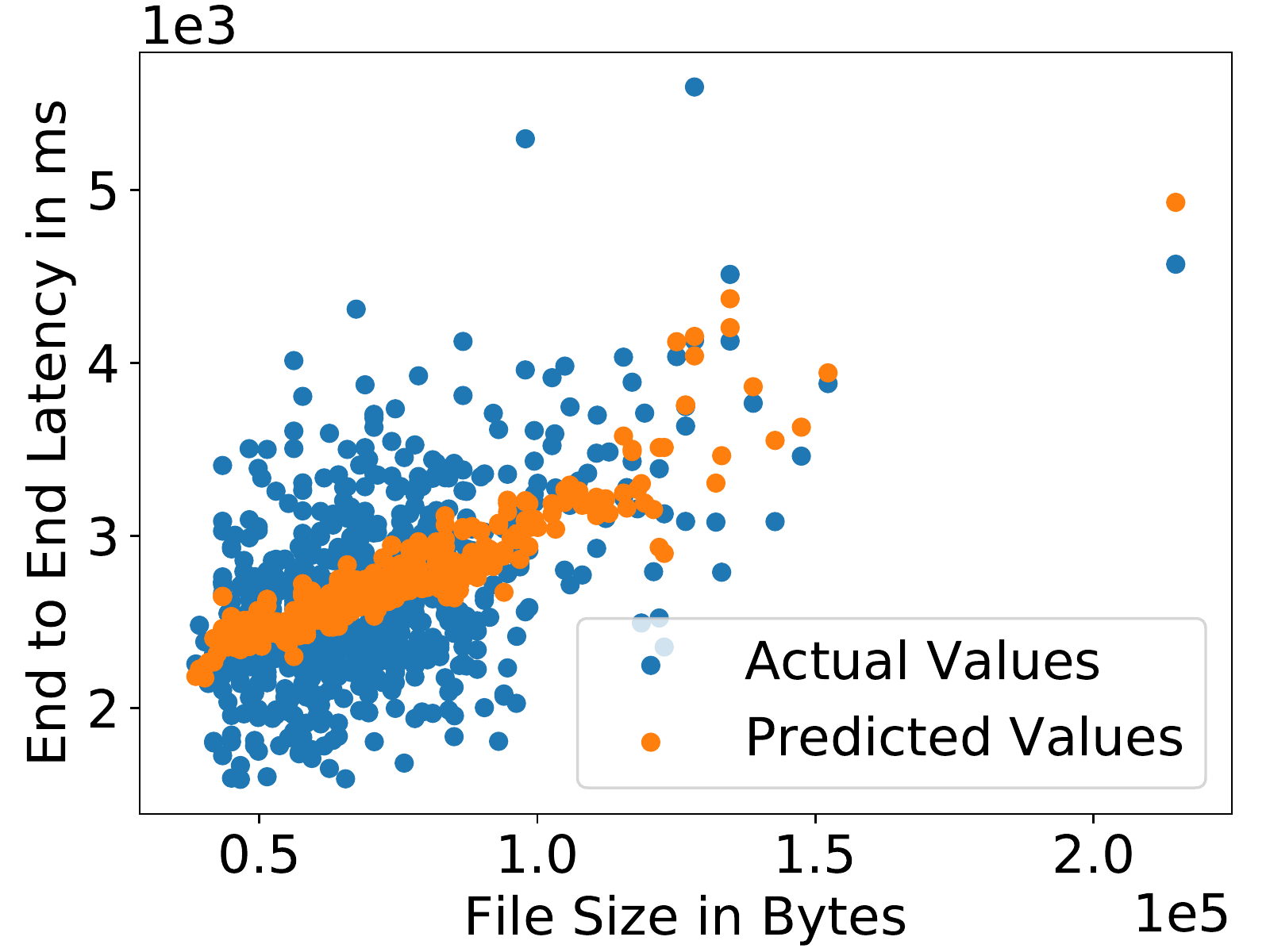}
      \caption{STT.}
      \label{fig.cloud_test_stt}
   \end{subfigure}
   \caption{Performance of end-to-end model on test data set for 1536~MB cloud lambda memory configuration (warm starts).}
   \label{fig.cloud_test_perf_1536MB}
   \vspace{10pt}
%    \squeezeuppicture
\end{minipage}
\begin{minipage}{\columnwidth}
\centering
   \begin{subfigure}{0.45\linewidth}
      \centering
      \includegraphics[width=\linewidth]{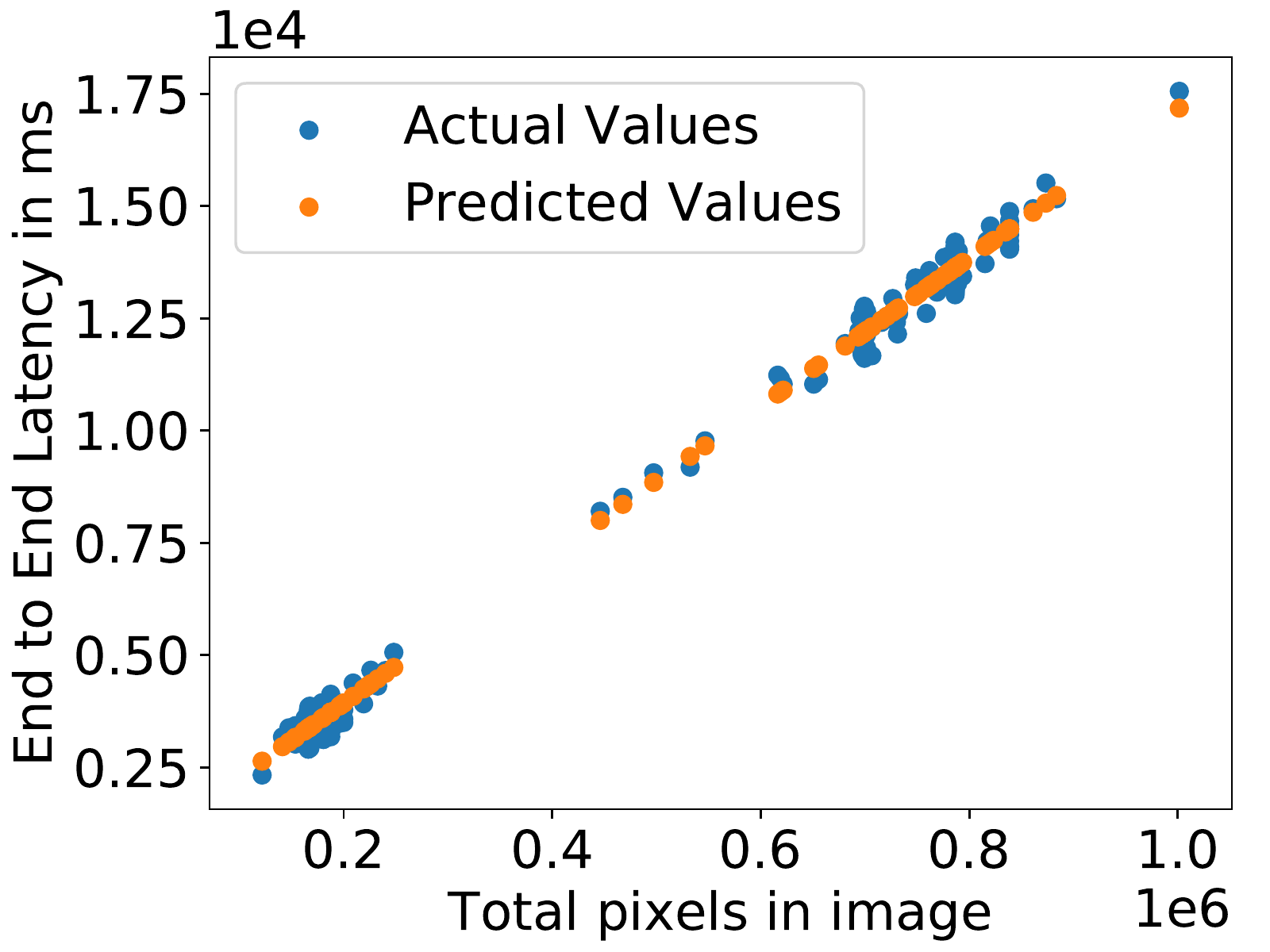}
      \caption{FD (MAPE: 3.78\%).}
      \label{fig.edge_test_facedetect}
   \end{subfigure}%
   \begin{subfigure}{0.45\linewidth}
      \centering
      \includegraphics[width=\linewidth]{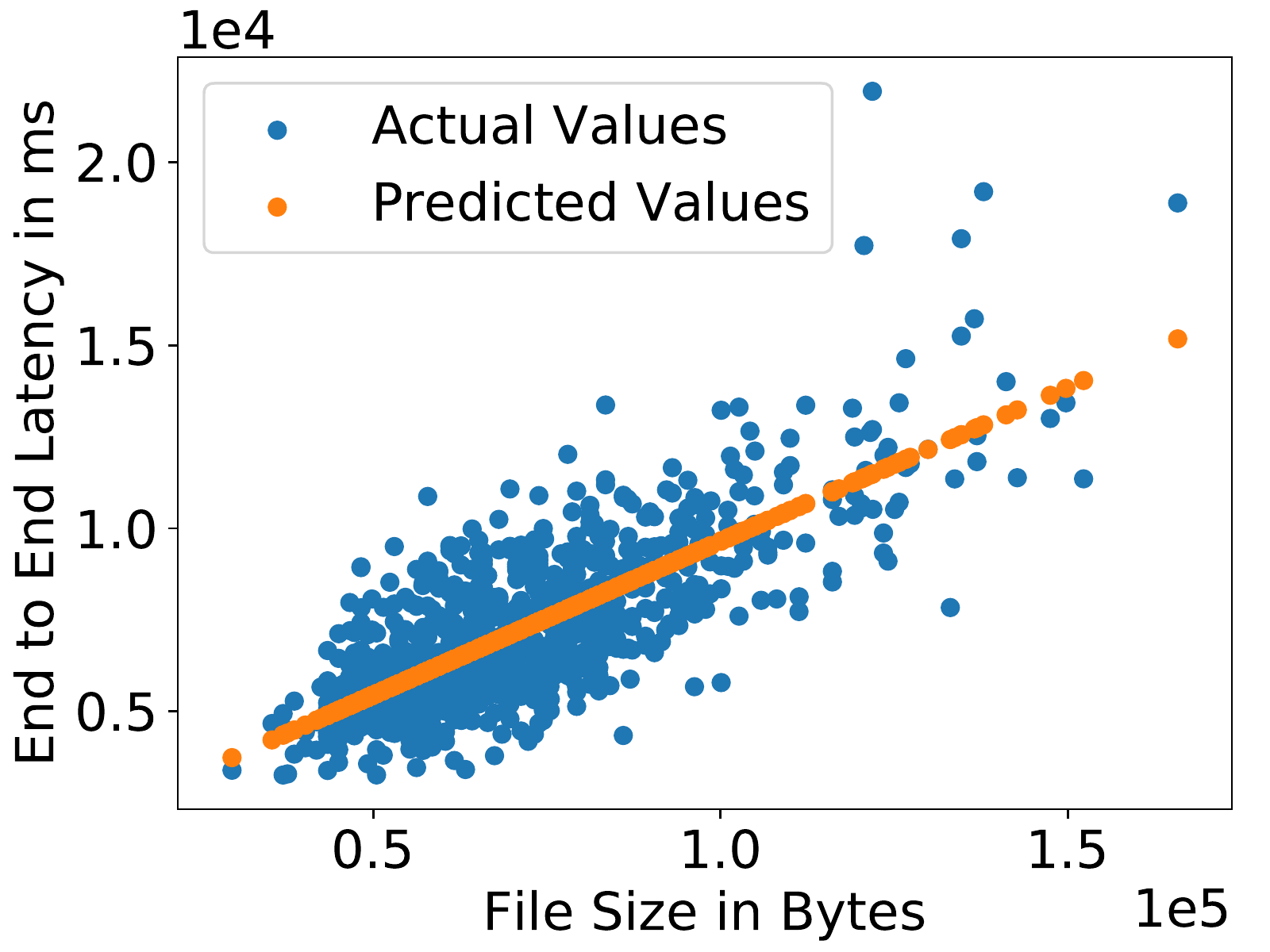}
      \caption{STT (MAPE: 15.70\%).}
      \label{fig.edge_test_speechtotext}
   \end{subfigure}
   \caption{Performance of edge end-to-end model on test dataset for edge pipelines.}
   \label{fig.edge_test_composite}
%    \squeezeuppicture 
\end{minipage}
\vspace{-15pt}
\end{figure}

In general, more variability in performance leads to higher MAPE values. Fig.~\ref{fig.cloud_test_perf_1536MB} depicts the end-to-end latency predicted by our model for the cloud pipeline for the FD and STT applications, with a  1536~MB warm start configuration. The figure also shows the measurements from the test data.  We can see that even for a single lambda function configuration and similar input file sizes, there is a notable variance in end-to-end latency, which presents challenges for generating accurate predictions.
 Fig.~\ref{fig.edge_test_composite} depicts the end-to-end latency prediction and the test data measurements for FD and STT in the edge pipelines.
 We note that the edge pipelines exhibit far less variation, and thus it is possible to predict the performance of these pipelines more accurately.

\begin{table}%[htbp]
\centering
\vspace{4pt}
\caption{Mean Absolute Percentage Error in the end-to-end latency of the cloud and edge pipelines.}
\footnotesize
\begin{tabular}{|l|c|c|c|} 
\hline
 & IR & FD & STT \\ 
\hline
Cloud & 25.38 & 13.24 & 14.56 \\ 
\hline
Edge & 2.15 & 3.78 & 15.70 \\
\hline
\end{tabular}
\label{table.mape}
\vspace{-5pt}
% \squeezeuppicture
\end{table}

\section{Framework Implementation} \label{sec.framework}
%In Sec.~\ref{sec.framework_overview}, we gave a brief overview of the framework. 
In this section, we elaborate on implementation details of the framework, specifically, the Predictor and the Decision Engine, and how they use the prediction models in Sec.~\ref{sec.perfmodel} to solve the optimization problems. %(\ref{subsub.Problem1}) and (\ref{opt:optim_time_cost_surplus}). 
\subsection{Predictor}
\sedit{Recall that the job of the Predictor is to predict the cost and end-to-end latency for a given input for every container configuration.}
Given the end-to-end latency prediction models of executions with cold and warm starts and the container idle time model, to predict the end-to-end latency, 
the  Predictor must determine whether a function execution will lead to a cold start or a warm start.
Since AWS does not expose any API for obtaining this information during execution, the Predictor maintains an offline data structure, the active container information list (CIL) that estimates which container configurations are warm in the AWS cloud. 
%In the cloud, when an `idle' container exists for a lambda function, invoking it will cause warm start. Else if all containers are busy or there exists no container, invoking it will cause cold start. Therefore, inside the offline data structure, \emph{Predictor} keeps track of the active containers by maintaining an active container information list (\emph{CIL}) for each cloud $\lambda_m$. 
For each $\lambda_m$, the CIL stores a list of active containers, and for each container, it keeps track of (1) whether the container is `idle' or `busy' executing a function, (2) the completion time of latest function executed within that container, and (3) the estimated time the container will be destroyed, obtained using (2) and $T_{idl}$.  %\sep{Is this the correct information?}
%(c) when was the last invocation time for this container.
%(c) estimated container lifetime (using $T_{idl}$). 

The Predictor exposes two methods to the Decision Engine: \texttt{predict}, that takes the input file as input and returns predictions of the end-to-end latencies and costs for $\lambda_m \in \Phi$ and for $\lambda_{edge}$. 
To generate these predictions, for each $\lambda_m \in \Phi$, the Predictor queries the CIL to determine if there exists an `idle' container for that $\lambda_m$.
If so, the Predictor generates the end-to-end latency using the model in Sec.~\ref{subsec.cloud-performance-model} with a warm start time. In case there are no containers or all containers are `busy' for $\lambda_m$, then the Predictor generates the cold start end-to-end latency. 
The Predictor uses the model in Sec.~\ref{subsec.edge-performance-model} to predict the end-to-end latency for $\lambda_{edge}$.
The costs for each configuration are then computed as described in Sec.~\ref{sub.pipelines}.  
%It is to be noted that since the predicted cost for a cloud configuration is based on predicted \cloudcomputetime, the cost is the same, irrespective of cold or warm start.

Once the Decision Engine has selected the configuration for the input, it invokes the \texttt{updateCIL} method on the Predictor, passing in the chosen configuration. The Predictor then updates the CIL, adding a new container if the lambda function execution results in a cold start, and updating the container status and function completion time, based on the estimated \cloudcomputetime\ or \edgecomputetime. If there are multiple `idle' containers for the selected configuration, we assume the function is assigned to the one with the most recent function completion time. 
%\sep{is it the most recent or the earliest?}. 
This assumption is based on our empirical observations of AWS Lambda.  In each call of \texttt{updateCIL}, the Predictor also checks for and removes dead containers from the CIL, based on estimated container lifetime information.

\begin{algorithm}[tbp]
    \caption{Minimize latency subject to cost constraint.}
    \label{algo.algorithm_min_time}
    \begin{algorithmic}[1] % The number tells where the line numbering should start
%        \Input
%        \Desc{}{Inputs, $\{\{\lambda_m, \forall m\}, \lambda_{edge}\}$, $\mathcal{C}_{max}, \alpha$}
%        \EndInput
        \Function{\underline{MinLatency} }{Inputs, $\Phi$, $\mathcal{C}_{max}, \alpha$} 
            \State $surplus = 0$ 
            \For{input $k$}           
				\State ($times$ , $costs$) := \Call{Predictor.predict}{input $k$}
				%\State \algComment{$\mathcal{T}$ contains end-to-end latencies of $\lambda_j \in \Phi \cup \{\lambda_{edge}\}$}
				%\State \algComment{$\mathcal{C}$ contains costs of $\lambda_j \in \Phi \cup \{\lambda_{edge}\}$}			
	           		 \State $\mathcal{M} := \{\lambda_j~|~\lambda_j \in \Phi \cup \{\lambda_{edge}\}~\text{and}$ \label{line.selectconfig_start} \\
			 ~~~~~~~~~~~~~~~~~~~ $costs(\lambda_j) \leq \mathcal{C}_{max} + \alpha \times surplus\}$ 
			        	\State  $config \gets \lambda_j \in \mathcal{M}$ with minimum latency \label{line.selectconfig_end}
			%	\If{$config = \lambda_{edge}$ and edge is busy and high queuing delay will be observed if queued at edge} 
		%			\State{$\mathcal{M} \gets \mathcal{M} \setminus \{\lambda_{edge}\}$}
			%		\If{$|M| \geq 1$}
				%		\State{$config \gets \lambda_j \in \mathcal{M}$ with minimum cost}
				%	\EndIf
				%\EndIf
		\State Use $config$ for function execution
                 \State $surplus += \mathcal{C}_{max} -costs(config)$
              	 \State \Call{Predictor.updateCIL}{$times(config),cost(config)$}
                %\State $\mathcal{C}_{avg} = \mathcal{C}_{avg} + \frac{\mathcal{C}(i)+ \mathcal{C}_{avg}}{n+1} $ 
            \EndFor
        \EndFunction
    \end{algorithmic}
\end{algorithm}

\subsection{Decision Engine}
The algorithm used by the Decision Engine for minimizing latency is described in Alg.~\ref{algo.algorithm_min_time}. The algorithm for minimizing cost is similar. The \emph{Decision Engine} obtains predictions from  the \emph{Predictor}.
%and uses  %Alg.~\ref{algo.algorithm_min_cost} and 
%Alg.~\ref{algo.algorithm_min_time} to solve the optimization problems for minimizing (\ref{opt:optim_time_cost_surplus}), \das{and similar algorithm for minimizing cost}. 
In both algorithms, for each input, the framework first uses \texttt{Predictor.Predict} to find the end-to-end latencies and costs for edge and cloud lambda functions.
If the objective is to minimize cost,  
%(Alg.~\ref{algo.algorithm_min_cost})
 the Decision Engine firsts create the list of configurations $\mathcal{M}$ that satisfies the latency constraint $\delta$. 
For each cloud configuration, the Decision Engine checks whether the predicted latency from the Predictor is less than $\delta$, and if so, adds the configuration to $\mathcal{M}$.
For $\lambda_{edge}$ the Decision Engine checks whether the predicted latency (from the Predictor) plus the predicted time in the Executor's FIFO queue (based on the predicted latency for any earlier tasks in the queue as well as any executing task) is less than $\delta$. If so, $\lambda_{edge}$ is added to $\mathcal{M}$.
 %as well as any executing edge function (based on the predicted latencies for these tasks).
 The Decision Engine then selects the configuration with the minimum predicted cost from $\mathcal{M}$. 
 If $\mathcal{M} = \emptyset$, there is no configuration that satisfies the deadline, so to save cost, the task is added to the Executor queue. 
 %If the Executor is busy and there exists $ \lambda_m \in \Phi$ that satisfies the performance requirements and the execution time including wait time in queue at edge is larger than the execution time at $\lambda_m$, the Decision Engine will choose that $\lambda_m$. Otherwise, the input will be added to the Executor queue.
 
% 
%\das{If $\lambda_{edge}$ is this selected configuration, and the Executor is busy and the edge execution time (including the queuing delay) is predicted to be very high, the algorithm checks for the next least expensive configuration in $\mathcal{M}$ that satisfies the deadline.} If no such configuration exists, the input is added to the Executor queue. 
%\sep{This next sentence needs to be updated for the change in edge queuing policy.}
%We note that adding the input to the queue may lead to a deadline violation \das{if  prediction is erroneous, however, with a reasonable deadline for an input, this allows us to execute more jobs at the edge.} 
%however, if the deadlines are chosen so that there is a cold start configuration that can meet the deadline for any input, then the input will never be queued at the edge.

If the objective is to minimize latency, the Decision Engine first creates the set of configurations $\mathcal{M}$ that satisfies the cost constraint, from the list of predictions returned by the Predictor.  It then selects the configuration with minimum predicted end-to-end latency from this set, where latencies are determined as described in the previous paragraph (lines \ref{line.selectconfig_start}-\ref{line.selectconfig_end} of Alg.~\ref{algo.algorithm_min_time}).
%\sep{This is confusing to me, since the line in the algorithm does not have anything to do with determining latencies. Maybe put the reference to the line earlier in the sentence?} \das{Changed it to include 5 - 7 because the whole thing happens between these 3 lines. Originally I gave reference to line 7 since the actual selection of a config was depicted in line 7. }. 
%\das{If the configuration with minimum latency is  $\lambda_{edge}$, and the Executor at the edge is busy and the predicted execution time including queuing time at edge is very high},
%the Decision Engine tries to find a cloud configuration in $\mathcal{M}$ that satisfies the cost constraint. If no such configuration exists, the input is placed in the Executor queue. Since the cost of execution $\lambda_{edge}$ is zero, this queuing will not result in a cost violation.
The Decision Engine then updates the surplus based on the predicted cost.
%Further, the Decision Engine can access the queue at the edge to verify if edge is busy or not. Since we implement the edge as non-blocking, we also add a logic in both algorithms that ensures if the edge is busy executing workload, but it is the best site in $\mathcal{M}$, then the execution is offloaded to the next best $\lambda_m$ in the cloud. 
At the end of both algorithms, the Decision Engine calls \texttt{Predictor.updateCIL} to update the CIL with container information. 
\section{Experiments} \label{sec.setup_results}
We first present an evaluation of our framework using simulation-based experiments, with measurements collected in AWS Lambda and Greengrass.
We then show results from a live experiment with our framework prototype.
%that solves problem (\ref{opt:optim_time_cost_surplus}) for the FD application in a live setting. 

\subsection{Simulation-Based Experiments}

For our experiments, we consider the same set of 19 cloud container configurations and the same edge configuration as used for the model training in Sec.~\ref{sec.perfmodel}.

% How do we collect actual measurements/generate "actual" data?
\sedit{We first collect warm latency measurements for a new set of input data for each application in each container configurations, both cloud and edge, using the process described in Sec.~\ref{sec.model_exps}. We use 600 input files for each application.
Since it difficult to collect a large number of cold start latency measurements, we instead simulate the cold start time by randomly selecting samples from the best-fit distribution on cold-start values from our training data. Similarly, we simulate $T_{idl}$ by randomly selecting samples from a normal distribution fitted on our observed measurements of container lifetime.}

% Description of simulation framework.
\sedit{We implement an event-driven simulation framework, which contains complete implementations of the Predictor and the Decision Engine.
The Predictor uses the trained models described in Sec.~\ref{models.sec}.
We feed input into the framework at intervals generated with a Poisson process, with arrival interval rate of four files per second for IR and FD and one file every ten seconds for STT. The Decision Engine selects a configuration based on the predicted end-to-end latency and cost.
We then simulate execution using the actual end-to-end latency and actual costs from the measured data.}

We initially perform all experiments using the training data to identify \emph{configuration sets}.
We observe that with a candidate set of all possible configurations, only a few configurations are ever selected. 
%The framework does not use the rest of the lambda memory types to execute jobs. 
We thus create sets that contain only the configurations the framework selected for the training data.
%We observe that the framework performance depends on configuration set, and we quantify the behavior in our experiments.
Every configuration set contains $\lambda_{edge}$ by default; we only state the elements of $\lambda_m$ explicitly for brevity.

We present results of our simulations for both optimization problems for the three applications.

%-----------------------------------------------------TABLES--------------
%---------------------------------------------MIN COSt TABLE---------------------------
\begin{table}[t]
    \centering
    \caption{Simulation: minimizing cost subject to deadline constraint. All configuration sets also include $\lambda_{edge}$. 
    }%\label{fig:table_min_total_cost} \hfill
    \begin{subtable}{\linewidth}
        \centering
        \caption{IR: $\delta$ = 2.7s, Avg. actual end-to-end latency $\approx1.37s$. }
        \resizebox{\linewidth}{!}{
        \begin{tabular}{|l|l|l|l|l|l|}
        \hline
        \multicolumn{1}{|c|}{Configuration Set} & \begin{tabular}[c]{@{}l@{}}Total Actual\\ Cost (\$)\end{tabular} & \begin{tabular}[c]{@{}l@{}}Cost Prediction\\ Error \%\end{tabular} & \multicolumn{1}{c|}{\begin{tabular}[c]{@{}c@{}}\% Deadlines\\ Violated\end{tabular}} & \multicolumn{1}{c|}{\begin{tabular}[c]{@{}c@{}}Average\\ Violation (ms)\end{tabular}} \\ \hline
\textbf{640,1024,1152} & 0.00155841 & 8.54 & 0.83 & 1.38 \\ \hline
640,1024,1408 & 0.00156019 & 5.88 & 1 & 1.73 \\ \hline
640,896,1152,1280 & 0.00156681 & 8.57 & 1.17 & 3.12 \\ \hline
640,768,1152 & 0.00157790 & 9.68 & 0.83 & 5.67 \\ \hline
\end{tabular}
%number of edge execs = 127
        \label{table.mincost_IR}}
    \end{subtable} \\
    \vspace{.2cm}
    \begin{subtable}{\linewidth}
        \centering
        \caption{FD: $\delta$ = 4.5s, Avg. actual end-to-end latency $\approx 2.43s$.}
        \resizebox{\linewidth}{!}{\begin{tabular}{|l|l|l|l|l|l|}
        \hline
        \multicolumn{1}{|c|}{Configuration Set} & \begin{tabular}[c]{@{}l@{}}Total Actual\\ Cost (\$)\end{tabular} & \begin{tabular}[c]{@{}l@{}}Cost Prediction\\  Error \%\end{tabular} & \multicolumn{1}{c|}{\begin{tabular}[c]{@{}c@{}}\% Deadlines\\ Violated\end{tabular}} & \multicolumn{1}{c|}{\begin{tabular}[c]{@{}c@{}}Average\\ Violation (ms)\end{tabular}} \\ \hline
\textbf{1280,1408,1664} & 0.01470774 & 0.26 & 0.33 & 3.7 \\ \hline
1152,1408,1664 & 0.01475062 & 0.49 & 0.33 & 3.27 \\ \hline
1152,1536,1792 & 0.01483715 & 2.85 & 0.5 & 1.72 \\ \hline
1280,1408,1536,1792 & 0.01483860 & 3.38 & 0.67 & 4.25 \\ \hline
%1152,1536,1664 & 0.0148738 & 3.2 & 0.83 & 2.39 \\ \hline
\end{tabular}
%number of edge execs = 33
        \label{table.mincost_FD}}
    \end{subtable}  \\
    \vspace{.2cm}
    \begin{subtable}{\linewidth}
        \centering
        \caption{STT: $\delta$ = 5.5s, Avg. actual end-to-end latency $\approx 3.35s$.}
        \resizebox{\linewidth}{!}{\begin{tabular}{|l|l|l|l|l|l|}
        \hline
        \multicolumn{1}{|c|}{Configuration Set} & \begin{tabular}[c]{@{}l@{}}Total Actual\\ Cost (\$)\end{tabular} & \begin{tabular}[c]{@{}l@{}}Cost Prediction\\ Error \%\end{tabular} & \multicolumn{1}{c|}{\begin{tabular}[c]{@{}c@{}}\% Deadlines\\ Violated\end{tabular}} & \multicolumn{1}{c|}{\begin{tabular}[c]{@{}c@{}}Average\\ Violation (ms)\end{tabular}} \\ \hline
\textbf{768,1152,1280,1664} & 0.019970506 & 2.49 & 6.17 & 49.6 \\ \hline
\begin{tabular}[c]{@{}l@{}}640,768,1280,\\ 1664,1792\end{tabular} & 0.020009885 & 1.75 & 7.83 & 71.94 \\ \hline
\begin{tabular}[c]{@{}l@{}}640,768,896,\\ 1280,1664\end{tabular} & 0.020022751 & 1.91 & 7.67 & 66.49 \\ \hline
%All & 0.0199976 & 3.22 & 7.67 & 67.86 \\ \hline ||||| REMOVED
640,896,1152,1664 & 0.020223292 & 3.33 & 6 & 58.40 \\ \hline
\end{tabular}
%number of edge execs = 58
        \label{table.mincost_STT}}
    \end{subtable}
    \squeezeuppicture
    \label{table.costvsdelta}
\end{table}
%---------------------------------------------MIN LATENCY TABLE---------------------------
\begin{table}[t]
    \centering
    \caption{Simulation: minimizing latency subject to cost constraint. All configuration sets also include $\lambda_{edge}$. }%\label{fig:table_min_end-to-end-latency}
    \begin{subtable}{\linewidth}
    \centering
\caption{IR: $\Cmax=\$5.33442\times10^{-06}, \alpha=0.02$.}% (actual end-to-end latency $>$ optimal by 47\% on avg.)}
\resizebox{\linewidth}{!}{
\begin{tabular}{|l|l|l|l|l|l|}
\hline
\multicolumn{1}{|c|}{Configurations} & \multicolumn{1}{c|}{\begin{tabular}[c]{@{}c@{}}Avg. Actual\\ Time/Task (s)\end{tabular}} & \multicolumn{1}{c|}{\begin{tabular}[c]{@{}c@{}}Latency Prediction\\  Error \%\end{tabular}} & \multicolumn{1}{c|}{\begin{tabular}[c]{@{}c@{}}\% Constraints\\ Violated\end{tabular}} & \begin{tabular}[c]{@{}l@{}}\% Budget\\  Used\end{tabular} \\ \hline
\textbf{1408,1664,2944} & 1.30 & 9.72 & 2.33 & 84.8 \\ \hline
1536,1664,2048,2944 & 1.314 & 7.90 & 2.17 & 88.6 \\ \hline
1280,1536,1664,2944 & 1.315 & 7.99 & 2.17 & 88.7 \\ \hline
1280,1408,1536,2944 & 1.329 & 10.73 & 1.83 & 84.8 \\ \hline
\end{tabular}
\label{table.thumbnailtimings}}
    \end{subtable} \\
    \vspace{.2cm}
    \begin{subtable}{\linewidth}
        \centering
\caption{FD: $\Cmax= \$2.96997\times10^{-05}, \alpha=0.02$.}%  (actual end-to-end latency $>$ optimal by 16.6\% on avg.)}
\resizebox{\linewidth}{!}{
\begin{tabular}{|l|l|l|l|l|l|}
\hline
\multicolumn{1}{|c|}{Configurations} & \multicolumn{1}{c|}{\begin{tabular}[c]{@{}c@{}}Avg. Actual\\ Time/Task (s)\end{tabular}} & \multicolumn{1}{c|}{\begin{tabular}[c]{@{}c@{}}Latency Prediction\\  Error \% \end{tabular}} & \multicolumn{1}{c|}{\begin{tabular}[c]{@{}c@{}}\% Constraints \\ Violated\end{tabular}} & \begin{tabular}[c]{@{}l@{}}\% Budget \\  Used\end{tabular} \\ \hline
\textbf{1536,1664,2048} & 2.1218 & 0.34 & 2.5 & 90.8 \\ \hline
1664,1920,2048 & 2.122 & 1.14 & 2 & 92.3 \\ \hline
1280,1664,2048 & 2.126 & 0.3 & 2.17 & 90.7 \\ \hline
1536,1664,1920 & 2.151 & 1.22 & 1.33 & 90.3 \\ \hline
%1536,2048 & 2.154 & 1.32 & 2.17 & 94.4 \\ \hline
\end{tabular}
\label{table.facedetecttimings}}
    \end{subtable}\\
    \vspace{.2cm}
    \begin{subtable}{\linewidth}
        \centering
        \caption{STT: $\Cmax=\$3.0747\times10^{-05}, \alpha=0.03$.}% (actual end-to-end latency $>$ optimal by 16.7\% on avg.)}
\resizebox{\linewidth}{!}{\begin{tabular}{|l|l|l|l|l|l|}
\hline
\multicolumn{1}{|c|}{Configurations} & \multicolumn{1}{c|}{\begin{tabular}[c]{@{}c@{}}Avg. Actual\\ Time/Task (s)\end{tabular}} & \multicolumn{1}{c|}{\begin{tabular}[c]{@{}c@{}}Latency Prediction\\  Error \%\end{tabular}} & \multicolumn{1}{c|}{\begin{tabular}[c]{@{}c@{}}\% Constraints \\ Violated\end{tabular}} & \begin{tabular}[c]{@{}l@{}}\% Budget \\  Used\end{tabular} \\ \hline
\textbf{1152,1280,1664} & 3.492 & 0.47 & 15.5 & 99.4 \\ \hline
1664 & 3.494 & 0.86 & 13.33 & 99.2 \\ \hline
1024,1280,1664 & 3.504 & 0.50 & 14 & 99.3 \\ \hline
1024,1152,1280,1664 & 3.561 & 0.85 & 15.17 & 99.3 \\ \hline
%All & 3.719 & 1.30 & 21.5 & 98.8 \\ \hline  ||||| REMOVED
\end{tabular}
\label{table.stttimings}}
    \end{subtable}
    \label{table.timevsalpha}
\vspace{-4mm}
\end{table}

\begin{figure*}[t]
\centering
	\begin{subfigure}{0.29\linewidth}
      \centering
      \includegraphics[width=\linewidth]{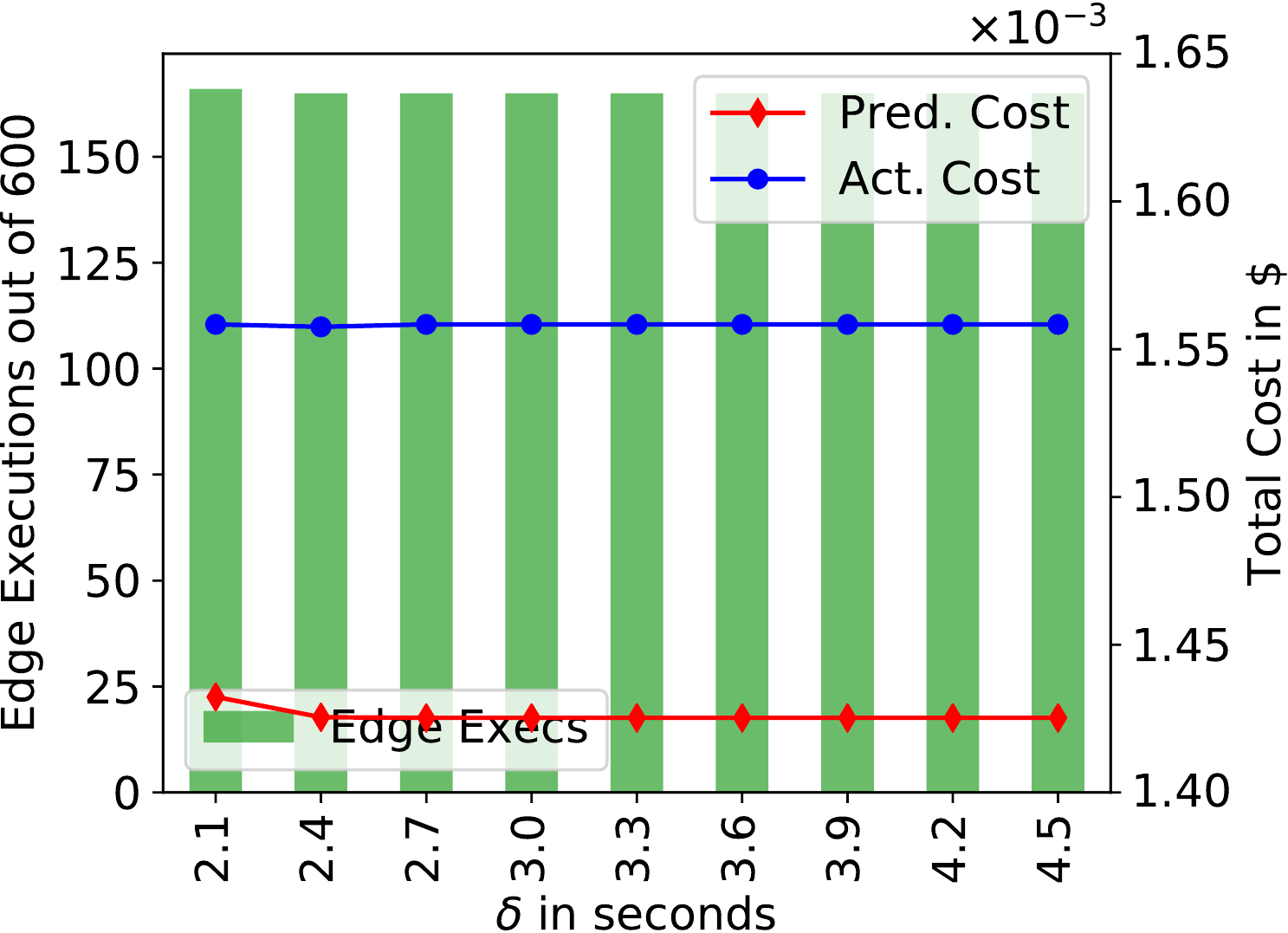}
      \caption{Image Resizing.}
      \label{fig.tncost}
   \end{subfigure}~~~~%	
	\begin{subfigure}{0.29\linewidth}     
      \centering
      \includegraphics[width=\linewidth]{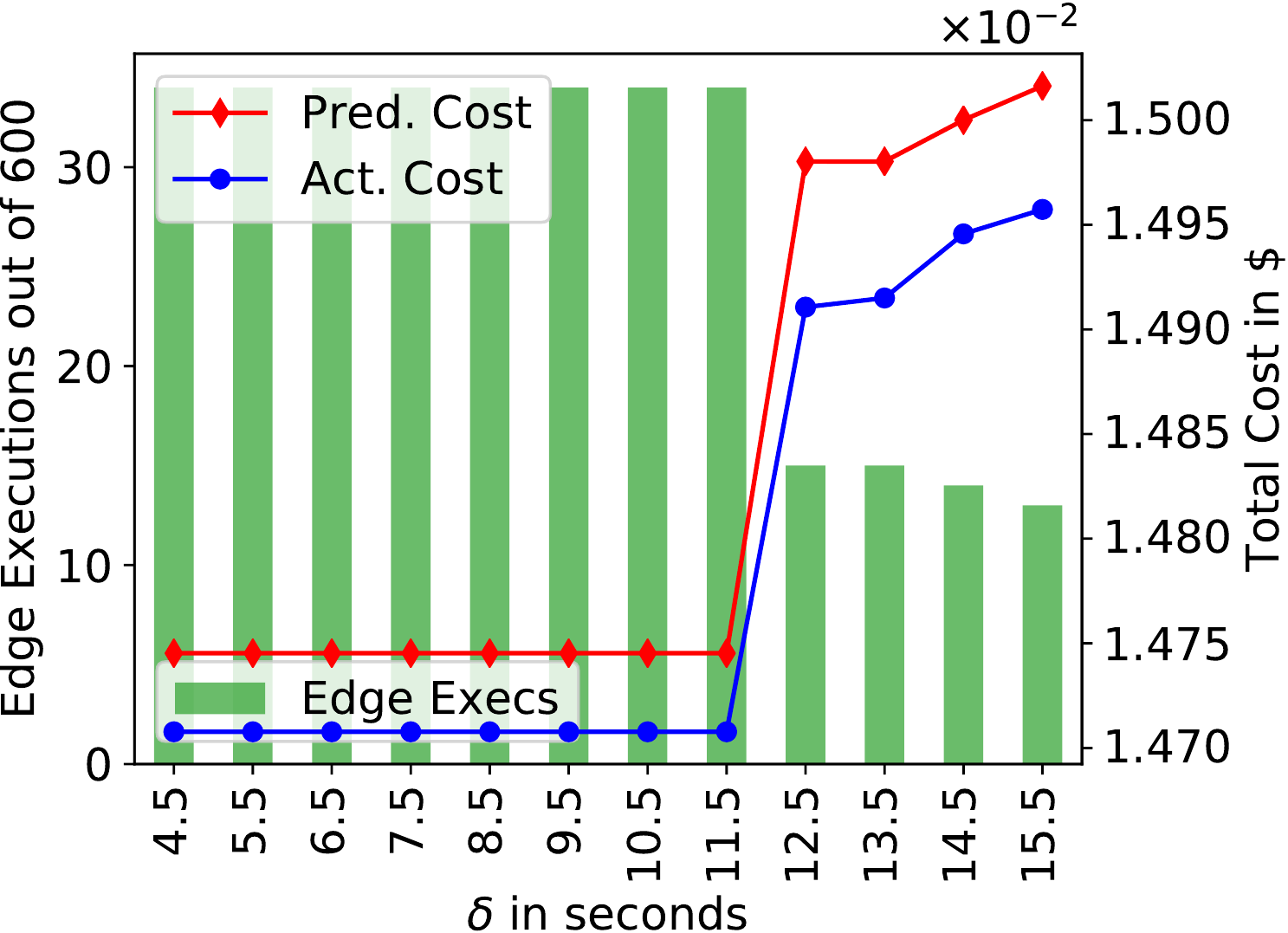}
      \caption{Face Detection.}
      \label{fig.fdcost}
   \end{subfigure}~~~~%
   \begin{subfigure}{0.29\linewidth}
      \centering
      \includegraphics[width=\linewidth]{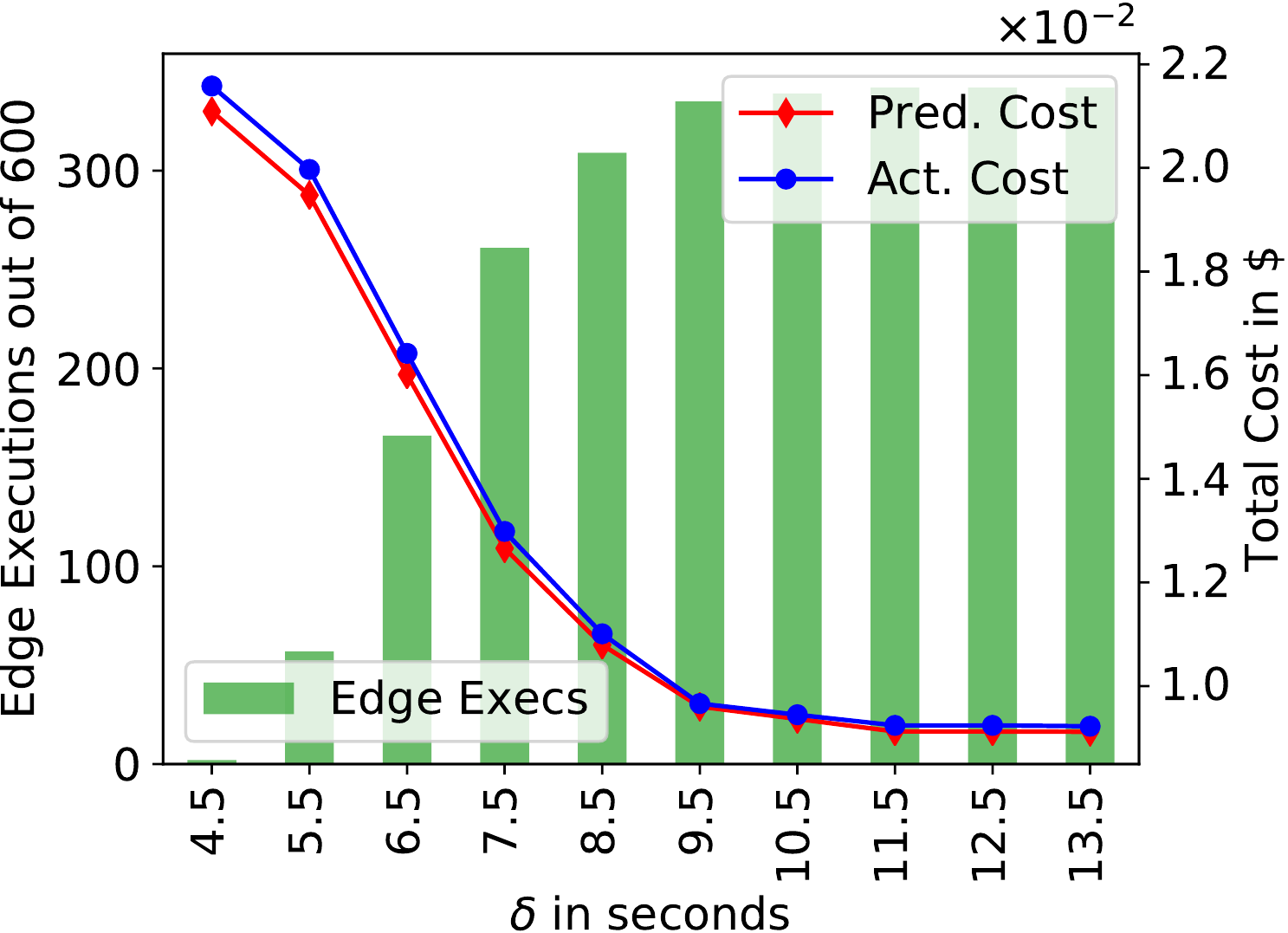}
      \caption{Speech-To-Text.}
      \label{fig.sttcost}
   \end{subfigure}
   \caption{Total execution cost (right Y axis) in \$ vs. $\delta$ (in seconds) for best performing configuration of different applications in minimizing total cost. The bar chart (left Y axis) represents number of edge executions out of 600.}
   \label{fig.costvsdelta}
   \squeezeuppicture
\end{figure*}

%---------------------------------------------MIN LATENCY FIGURE---------------------------
\begin{figure*}[t]
\centering
	\begin{subfigure}{0.29\linewidth}
      \centering
      \includegraphics[width=\linewidth]{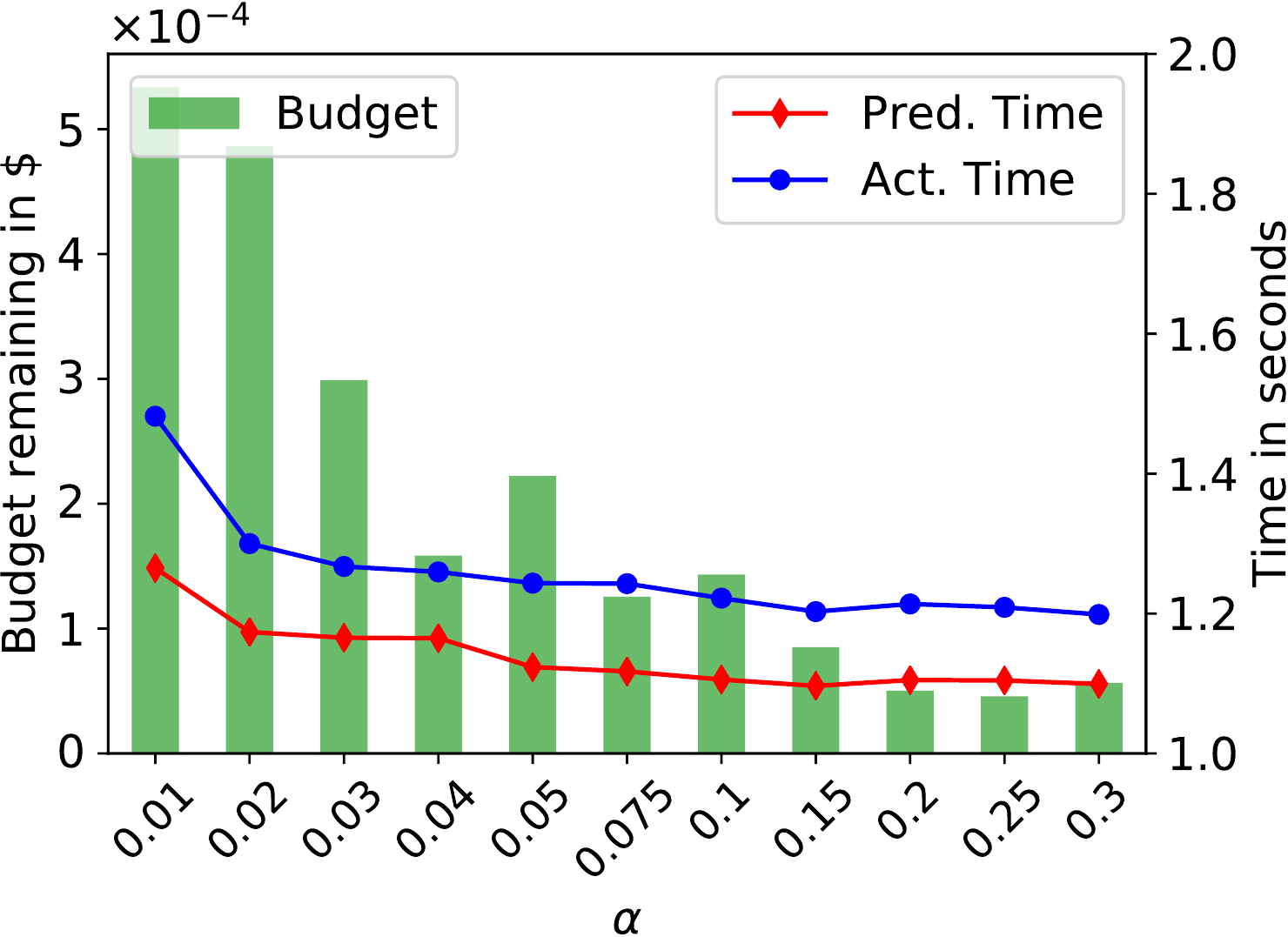}
      \caption{IR: $\Cmax=5.334 \times 10^{-05}\$$.}
      \label{fig.tncost}
   \end{subfigure}~~~~%
	\begin{subfigure}{0.29\linewidth}     
      \centering
      \includegraphics[width=\linewidth]{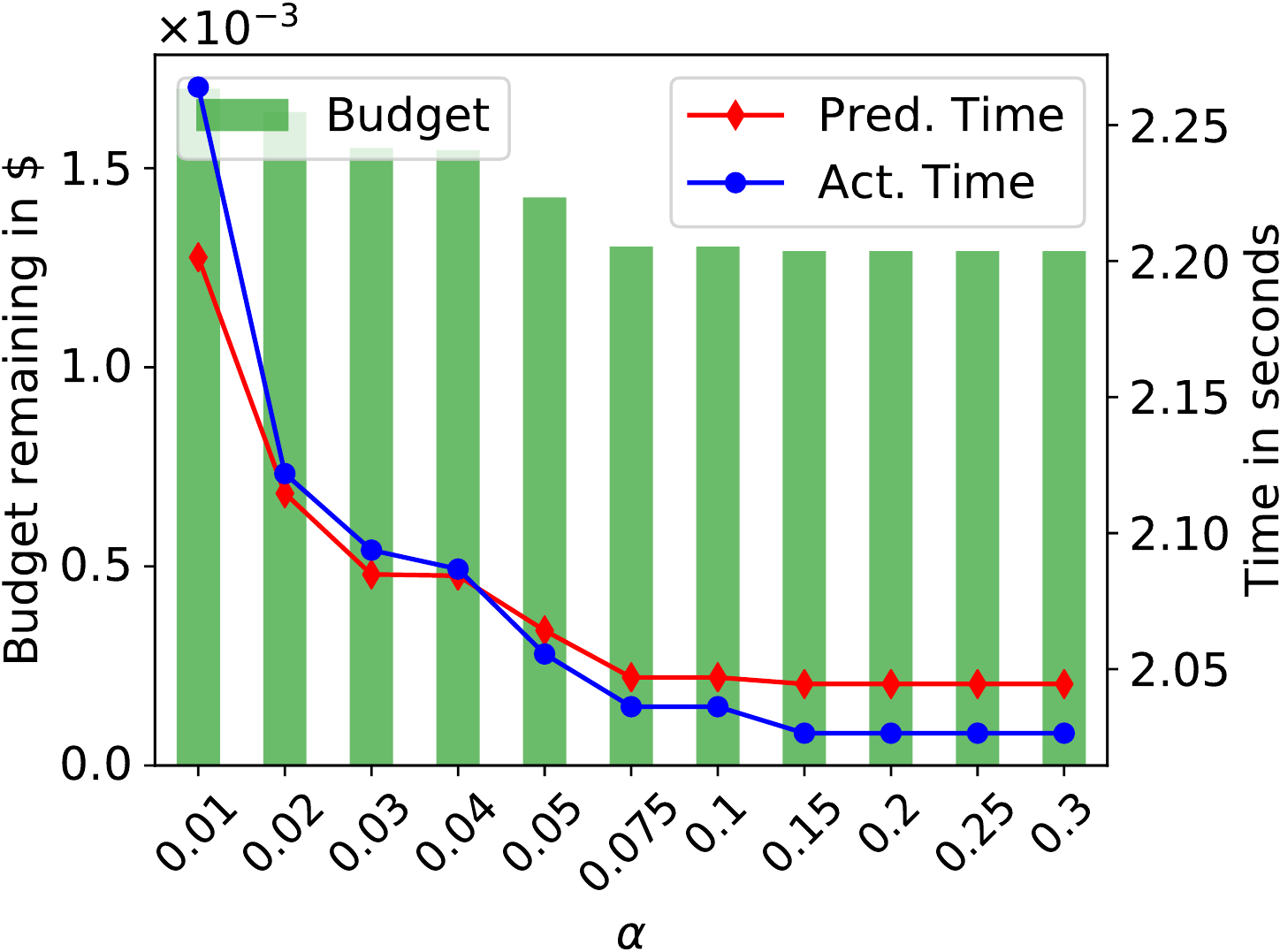}
      \caption{FD: $\Cmax=2.97 \times 10^{-05}\$$.}
      \label{fig.fdcost}
   \end{subfigure}~~~~%
   \begin{subfigure}{0.29\linewidth}
      \centering
      \includegraphics[width=\linewidth]{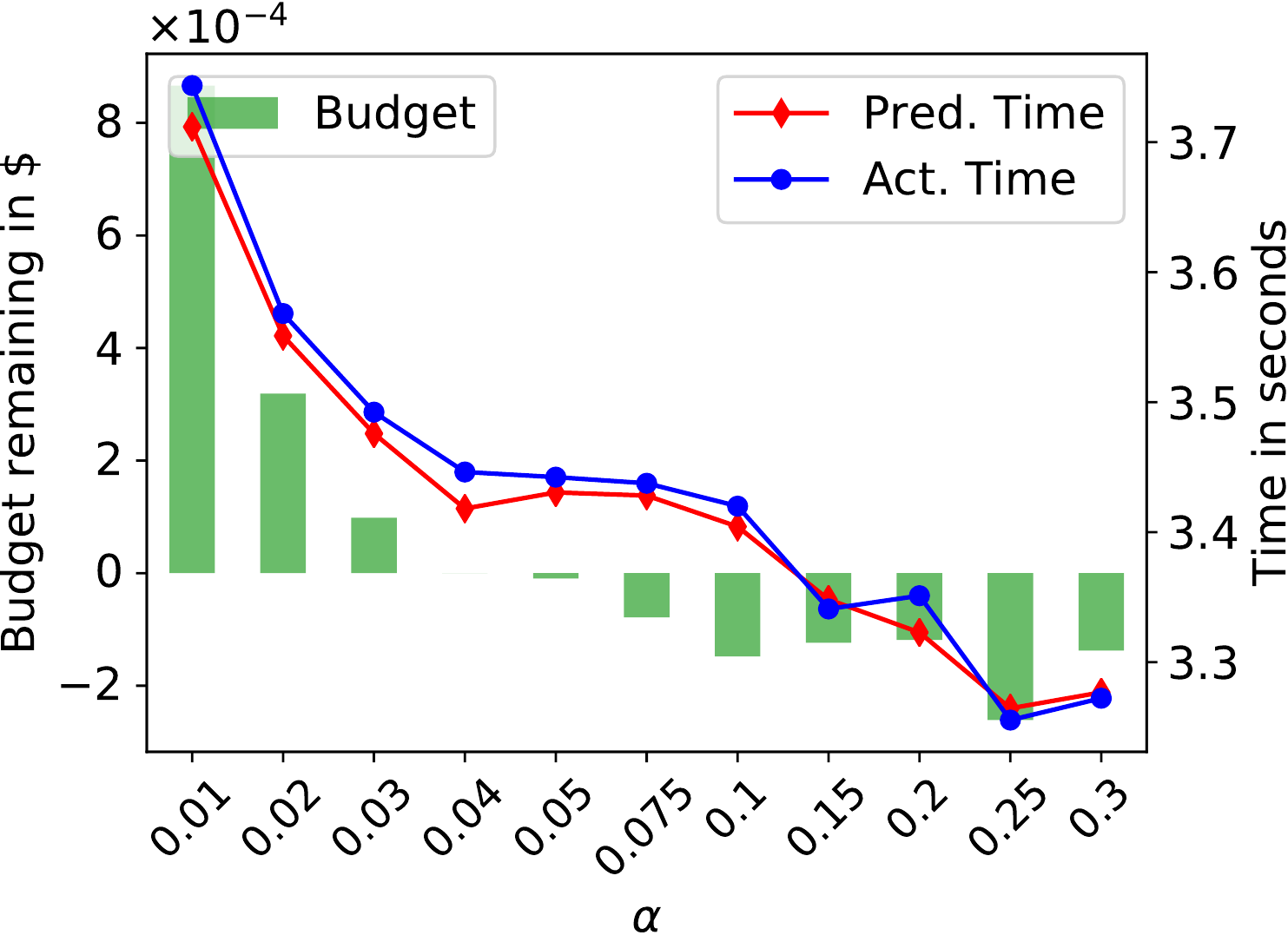}
      \caption{STT: $\Cmax=3.074 \times 10^{-05}\$$.}
      \label{fig.sttcost}
   \end{subfigure}   
   \caption{Average end-to-end latency (right Y axis) vs. $\alpha$ for best performing configuration of different applications in minimizing end-to-end latency. The bar chart (left Y axis) represents total budget \$ remaining at the end of execution.}
   \label{fig.timevsalpha}
   \squeezeuppicture
   \vspace{-3mm}
\end{figure*}
%----------------------------------------------------------------------------------------

\subsubsection{Cost Minimization} \label{subsub.min_total_cost}

We first evaluate our solution for  cost minimization subject to a per-function-execution deadline.
We select the deadline $\delta$ for each application based on the training data, ensuring that each configuration set contains a feasible configuration for every input in the training set.

In Table~\ref{table.costvsdelta}, we present the performance of different configuration sets in increasing order of total actual cost for each application, along with the percentage error between the actual and predicted total cost. The total costs are computed over all 600 inputs. We measure cost prediction error \% as the absolute percentage error between the total actual cost and total predicted cost. We also show the percentage of inputs where actual end-to-end latency violated the deadline.

We observe that the configuration sets {$\{$640MB, 1024MB, 1152MB$\}$}, {$\{$1280MB,1408MB,1664MB$\}$}, and $\{$768MB, 1152MB, 1280MB, 1664MB$\}$ achieve the smallest total actual cost in the IR, FD, and STT applications, respectively. We also observe that a smaller cost prediction error leads to better performance in terms of the total cost.
In general, smaller function execution times are prone to higher cost prediction error.  AWS quantizes billed amount in multiples of 100ms,  e.g., 98 ms compute time would be rounded to 100ms, whereas a 101ms compute time will be rounded to 200ms,  and so a small error in the execution time prediction may result in a larger error in cost prediction when the magnitude of execution time is low. 
We also see that fewer deadline violations are correlated with better performance in terms of minimizing total cost.

In Fig.~\ref{fig.costvsdelta}, we plot the predicted and actual total costs versus the deadline $\delta$ for the best configuration (in bold) in Table~\ref{table.costvsdelta} for each application. 
\sedit{For all three applications, our predicted cost closely mirrors the actual cost.}
We observe that for IR, the number of edge executions does not appear to be correlated with the deadline. This is because for IR, in general, the edge pipeline execution is faster than the cloud pipeline execution. 
We also observe that for FD, as the deadline increases, the number of edge executions decreases, and accordingly, the cost increases. This is because, with a larger deadline, tasks are assigned to the edge, causing the edge to be busy, which in turn, leads to more tasks being assigned to the cloud.
STT exhibits a more expected behavior; as the deadline increases, more tasks are executed at the edge. This is in part due to the slower input rate for STT, which increases the availability of the edge for task execution.

We observe that the absolute error between predicted and actual total cost for the best performing configurations of FD and STT are less than 3\%. Also, all other configurations for FD and STT performs well, with less than 4\% absolute total cost prediction error.
Finally, we observe that warm start vs. cold start prediction mismatches can influence the total cost prediction error. A slower input rate reduces the chances of mis-predicting cold and warm starts; STT has 0\% absolute error between estimated and actual cold and warm starts, while for FD it is 2.5\%.

\subsubsection{Latency Minimization} \label{subsub.min_avg_time_st_cost}
We next evaluate our framework in solving the problem of minimizing latency subject to a task cost constraint.
We use the formulation in Eqn. (\ref{opt:optim_time_cost_surplus}), which allows surplus budget to be spent on subsequent tasks.
For each application, we select $\Cmax$ and $\alpha$ from experiments on the training data set. We select $\Cmax$ and $\alpha$ to be small enough so that for some inputs, it is necessary to use $\lambda_{edge}$.
Similar to the previous experiments, we select configuration sets that consist of configurations that the framework selected when processing the training data.

We measure latency prediction percentage error as the absolute percentage error between actual and predicted average end-to-end latency at the end of the simulation. Further, we measure the percentage of constraints violated as the percentage of tasks where the actual cost of execution violated the corresponding cost constraint.
%\sep{Explain how latency prediction error and \% deadlines violated are computed.}
The percentage of budget used is computed as the total actual cost for processing the input workload divided by the total budget for the input workload, $\Cmax \times \text{number of inputs}$.

In Table~\ref{table.timevsalpha}, we present the results for different configuration sets in increasing order of average end-to-end latency.
The table also shows the latency prediction error, percentage of function executions that violated the cost constraints, and the percentage of total budget used.
We observe that configurations $\{$1408MB, 1664MB, 2944MB$\}$, $\{$1536MB,1664MB, 2048MB$\}$, and $\{$1152MB,1280MB,1664MB$\}$ achieve the minimum end-to-end latency for the IR, FD and STT applications, respectively. We also observe that these configurations have low latency prediction error (except IR due to its high variance). Further, even though the cost constraints were violated for some inputs, the total cost of execution of the entire input workload was always under the total budget.

%\sep{Can you add a statement about the accuracy of the framework to this paragraph (based on the figure)?}
In Fig.~\ref{fig.timevsalpha}, we plot the actual and predicted average end-to-end latency for various values of $\alpha$ with $\Cmax$ fixed. We use the best configuration set per application, shown in bold in Table~\ref{table.timevsalpha}. We observe that the actual average  latency obtained from the framework execution closely follows the predicted average latency, with less than 2\% absolute error for FD and STT and less than 11\% error for IR. 
%Though the absolute prediction error of the best configuration of the applications is less than 5\%.

We observe that in all applications, with increasing $\alpha$ the average end-to-end latency decreases. By increasing $\alpha$, more surplus budget is available per task, and thus, more cloud configurations can satisfy the cost constraint. 
These cloud configurations typically have shorter executions times. We further observe that for FD, the total remaining budget does not vary much with $\alpha$, and in both IR and STT, the budget remaining decreases with increased $\alpha$. 
%Higher values of $\alpha$ essentially means using up the budget surplus in the next few jobs. 
With a smaller $\Cmax$, a larger value of $\alpha$ may lead to budget violations. For example, in STT, we observe that for the best configuration $\{$1152MB,1280MB,1664MB$\}$, as we increase $\alpha$ above $0.04$, the total budget remaining becomes negative, which means the total actual cost went over the total budget.
%The schedules produced here would be deemed `infeasible'. 
Also, for $\alpha=0$, we observe very high average end-to-end latencies: IR $=10.5\,s$, FD $=452.2\,s$, and STT $=12.64\,s$. This is due to the fact many tasks are run on the edge, as the cost constraint restricts cloud executions. As a result, the waiting periods in the edge queue leads to an increase the average execution time.

\subsection{Live Evaluation}

To demonstrate the effectiveness of our framework in a real-world application, we have implemented a prototype of our framework.
We evaluate this prototype in AWS Greengrass and Lambda using the FD application, with the same 600 input files used in the simulations.
The framework is configured to minimize end-to-end latency subject to a cost constraint.
We use the edge configuration described in Sec.~\ref{sec.model_exps}. For the cloud, we use the best configuration set from the simulations,
$\{$1536MB, 1664MB, 2048MB$\}$.

We measure the accuracy between the predicted and actual latency.  Further, we measure how many times the framework violates the budget and the percentage of the budget remaining at the end of workload. % among the inputs. 
We also measure the number of times we mis-predict `cold' or `warm' starts.
We perform the experiment four times and show the average results. 

\begin{table}[htb]
\vspace{-3pt}
\centering
\caption{Average results over four runs of the FD application with configuration set $\{$1536MB, 1664MB, 2048MB$\}$, $\Cmax = 2.96697\times 10^{-05}$, and $\alpha = 0.02$.}
\resizebox{\linewidth}{!}{
\begin{tabular}{|l|l|l|l|l|}
\hline
 \begin{tabular}[c]{@{}l@{}}Avg. Actual \\ End-To-End\\ Latency\end{tabular} & 
 \begin{tabular}[c]{@{}l@{}}Latency \\ Prediction \\ Error \end{tabular} & 
  \begin{tabular}[c]{@{}l@{}}Violations\\ of cost \\ budget\end{tabular}  &
 \begin{tabular}[c]{@{}l@{}} \% Budget \\ Used \end{tabular} &
 \begin{tabular}[c]{@{}l@{}} Warm-Cold \\ Mismatches \end{tabular} \\
 \hline
 1.71 s                                                                  & 
 5.65 \%                                                                             & 
 \begin{tabular}[c]{@{}l@{}}8 / 600\\ = 1.33 \%\end{tabular}           & 
 86 \%                                  &
\begin{tabular}[c]{@{}l@{}}5 / 600\\ = 0.83 \%\end{tabular}                      
                          \\ \hline
\end{tabular}
}\label{table.liveresults}
% \squeezeuppicture
\vspace{-2pt}
\end{table}

We present the results in Table~\ref{table.liveresults}. 
%The average actual end-to-end latency in the live experiments was 1.71s, whereas during simulation it was around 2.12s \sep{Can you provide any insight as to why this is the case? If not, I would remove this sentence}. 
\sedit{Our latency prediction error is 5.65\%. While this is larger than the 0.34\% prediction error observed in simulations, the prediction accuracy is still quite high.
Also, we find that the total actual cost is under the total budget, with $\approx 86\%$ of the total budget used. The warm start/cold start prediction error is also low, at 5 mis-predictions out of 600 inputs. These results suggest that our framework works well in practice.}
Finally, we note that when the same input workload is processed only using the edge pipeline, the average end-to-end latency is 2404\,s due to queuing and is impractical compared to 1.71\,s with cloud offload. 
\section{Related Work} \label{sec.relatedwork}
Various approaches for task placement and computation offloading have been proposed in the context of mobile cloud computing in recent years
using static program analysis and annotations~\cite{maui,clonecloud,shi2014cosmos}, as well as data flow graph-based dynamic partitioning~\cite{odessa}.
%MAUI~\cite{maui}, CloneCloud~\cite{clonecloud}, Cosmos~\cite{shi2014cosmos} uses either static user annotations or program analysis, whereas Odessa~\cite{odessa} uses data flow graph based dynamic partitioning to decompose the user code into offloadable methods or tasks. 
%The offloading decision is then made by profiling the application execution latency, energy or bandwidth usage and user-defined performance objectives.
%\sep{Can you specify which papers use which of the two methods?}. 
In these works, tasks are offloaded from mobile devices to either VMs in the cloud or to remote servers. 
The approaches in \cite{maui, clonecloud} formulate the offloading problem as an ILP, whereas \cite{shi2014cosmos, odessa} depend on carefully designed greedy heuristics.
%More recently, offloading of neural network computation has been explored in NeuroSurgeon \cite{kang2017neurosurgeon} and IONN \cite{jeong2018ionn}. 
More recently, offloading of neural network computation has been explored~\cite{kang2017neurosurgeon,jeong2018ionn}. 
In these works, deep neural network layers are partitioned across edge devices and cloud servers and executed collaboratively to satisfy latency, accuracy, and energy-saving objectives.

% \das{The following commented sentence can be safely cut I believe as it is also mentioned in last paragraph. Also maybe we can reduce 2 citations here.}
%Many previous works have focused on offloading to cloud VMs or private servers.
%To maximize the performance of offloaded applications, 
Cloud performance optimization in context of VMs has been
%studied extensively through VM allocation~\cite{venkataraman:2016:eep:2930611.2930635,verma2011}, VM performance characterization~\cite{wang-varela-ucc-2011,awan2015}, and autoscaling~\cite{lorido-botran2014,roy2011efficient}.
studied extensively through VM allocation~\cite{venkataraman:2016:eep:2930611.2930635},
VM performance characterization~\cite{awan2015}, and autoscaling~\cite{lorido-botran2014,roy2011efficient}.
In contrast, we use serverless functions as the cloud offload destination. This imposes many behavioral constraints, for example, serverless is a stateless and event-based computation model, while VMs or servers are long-lived and stateful. 

Several recent works have studied performance characteristics of serverless systems across different industry platforms.
% such as AWS Lambda, IBM Openwhisk, Azure Functions, and Google Cloud Functions. 
The authors in \cite{serverlessicdcs} proposed their own serverless platform and compared its execution performance with industry platforms. Extensive studies has been done on scalability of platforms, function latency, infrastructure retention, infrastructure provisioning~\cite{lloyd2018serverless, wang2018peeking,figiela2018performance}, impact of language runtime on function performance~\cite{jackson2018investigation}, and latency of edge serverless platforms~\cite{edgebench}. \cite{spock} uses serverless functions to handle incoming workloads for the  duration it takes to allocate sufficient VMs to minimize SLA violations. 
The work \cite{kannan2019grandslam} tackles the problem of executing jobs in microservices under a SLA constraint from a platform provider's perspective by maximizing the utilization of provider hardware. In contrast, we study methods to reduce cost or end-to-end latency from a client's perspective in an edge-cloud system.  

Finally, the authors in Costless~\cite{costless} also present an algorithm that uses serverless functions for computational offloading from the edge. Their work however focuses on efficient partitioning of a chain of functions comprising an application, where some functions execute on the edge and some in the cloud, to reduce execution cost.
In contrast, we focus on data-driven predictive offload decision making, characterizing the effect of warm and cold starts, and lastly, on performance maximization with multiple objectives using different types of real-world applications. 
%\sep{I find the previous sentence confusing. Maybe it should be two sentences?}

%%%%%%%%%%%%%%%%%%%%%%%%%%%%%%%%%%%%%%%%%%%%%%%%%%%%%%%%%%%%%%%%%%%%%%%%%%%%%%%%
\section{Conclusion} \label{sec.conclusion}
We have presented a performance optimization framework for serverless applications in an edge-cloud platform.
As part of this framework, we have developed models for accurately predicting end-to-end latencies and cost for functions running in the cloud or the edge. 
We provide a simulation-based evaluation of our framework on three representative applications.
%, image resizing, face detection and speech-to-text. 
The best configurations achieved less than .3\% absolute cost prediction error when minimizing total cost and less than .4\% absolute latency prediction error when minimizing average latency.
We also present results of live experiments, run in AWS, using the face detection application.  Our evaluation shows that our framework can predict end-to-end latency with less than 6\% error and obtain almost three orders of magnitude average end-to-end latency minimization compared to a naive edge execution. 
In future work, 
%we plan to incorporate memory utilization and billing information from AWS logs into our prediction models, and 
we will expand our prediction methods to explicitly incorporate the high variance sometimes observed in serverless platforms.

%\subsubsection*{Future Work} In future we will incorporate memory usage and actual billing information from AWS logs into the offloading decision making process and explore more robust ways of improving predictions to reduce variance. 

%%%%%%%%%%%%%%%%%%%%%%%%%%%%%%%%%%%%%%%%%%%%%%%%%%%%%%%%%%%%%%%%%%%%%%%%%%%%%%%%
\section*{Acknowledgment}
This work is supported by the National Science Foundation under grants IIS 1900977, CNS 1553340 and CNS 1816307, and an AWS Cloud Credits for Research grant.
%more thanks here

%\addtolength{\textheight}{-3cm}   % This command serves to balance the column lengths
                                  % on the last page of the document manually. It shortens
                                  % the textheight of the last page by a suitable amount.
                                  % This command does not take effect until the next page
                                  % so it should come on the page before the last. Make
                                  % sure that you do not shorten the textheight too much.
                                  
%%%%%%%%%%%%%%%%%%%%%%%%%%%%%%%%%%%%%%%%%%%%%%%%%%%%%%%%%%%%%%%%%%%%%%%%%%%%%%%%

% trigger a \newpage just before the given reference
% number - used to balance the columns on the last page
% adjust value as needed - may need to be readjusted if
% the document is modified later
%\IEEEtriggeratref{8}
% The "triggered" command can be changed if desired:
%\IEEEtriggercmd{\enlargethispage{-5in}}

% references section

% can use a bibliography generated by BibTeX as a .bbl file
% BibTeX documentation can be easily obtained at:
% http://www.ctan.org/tex-archive/biblio/bibtex/contrib/doc/
% The IEEEtran BibTeX style support page is at:
% http://www.michaelshell.org/tex/ieeetran/bibtex/
%\bibliographystyle{IEEEtran}
% argument is your BibTeX string definitions and bibliography database(s)
%\bibliography{IEEEabrv,../bib/paper}
%
% <OR> manually copy in the resultant .bbl file
% set second argument of \begin to the number of references
% (used to reserve space for the reference number labels box)
\bibliographystyle{abbrv}
\bibliography{biblio} 
\end{document}

% that's all folks
\end{document}